\documentclass{aa}

\usepackage{graphicx}
\usepackage{txfonts}
\usepackage{pdflscape}
\usepackage{multicol}
\usepackage{lmodern}
\usepackage[usenames]{color}

\usepackage{natbib,twoopt}
\bibpunct{(}{)}{;}{a}{}{,} 

\bibpunct{(}{)}{;}{a}{}{,} 
\makeatletter
\newcommandtwoopt{\citeads}[3][][]{\href{http://adsabs.harvard.edu/abs/#3}%
{\def\hyper@linkstart##1##2{}%
\let\hyper@linkend\@empty\citealp[#1][#2]{#3}}}
\newcommandtwoopt{\citepads}[3][][]{\href{http://adsabs.harvard.edu/abs/#3}%
{\def\hyper@linkstart##1##2{}%
\let\hyper@linkend\@empty\citep[#1][#2]{#3}}}
\newcommandtwoopt{\citetads}[3][][]{\href{http://adsabs.harvard.edu/abs/#3}%
{\def\hyper@linkstart##1##2{}%
\let\hyper@linkend\@empty\citet[#1][#2]{#3}}}
\newcommandtwoopt{\citeyearads}[3][][]%
{\href{http://adsabs.harvard.edu/abs/#3}
{\def\hyper@linkstart##1##2{}%
\let\hyper@linkend\@empty\citeyear[#1][#2]{#3}}}
\newcommand\vv{{\mathrm v}  }
\newcommand{\Msun}{\ensuremath{\rm M_\odot}}

\makeatother

\begin{document}

\title{\bf Wind and nebula of the M\,33 variable GR\,290 (WR/LBV) \thanks{Based on observations made with the Gran Telescopio Canarias (GTC), installed at the Spanish Observatorio del Roque de los Muchachos of the Instituto de Astrofísica de Canarias, in the island of La Palma and with the Cassini 1.52-m telescope of the Bologna Observatory (Italy).}}
\titlerunning{Wind and nebula of GR\,290}

\author{Olga Maryeva \inst{1} \and Gloria Koenigsberger\inst{2} \and Oleg Egorov\inst{3} \and Corinne Rossi\inst{4}  
\and Vito Francesco Polcaro \inst{5}\fnmsep\thanks{Deceased on February 11, 2018}  \and Massimo Calabresi\inst{6} \and Roberto F. Viotti\inst{5}
}

\institute{Astronomical Institute, Czech Academy of Sciences, Fri\v{c}ova 298, 25165, Ond\v{r}ejov, Czech Republic, olga.maryeva@gmail.com \\
\and Instituto de Ciencias F\'{\i}sicas, Universidad Nacional Aut\'onoma de M\'exico, Ave. Universidad S/N,  62210, Cuernavaca, Morelos, M\'exico, gloria@astro.unam.mx \\
\and Lomonosov Moscow State University, Sternberg Astronomical Institute, Universitetsky pr. 13, 119234, Moscow, Russia, egorov@sai.msu.ru \\
\and Universit\`a Sapienza, Piazza A.Moro 5, 00185, Roma, Italy, corinne.rossi@uniroma1.it     \\
\and INAF-Istituto di Astrofisica e Planetologia Spaziali di Roma (IAPS-INAF), Via del Fosso del Cavaliere 100, 00133, Roma, Italy, \\
vitofrancesco.polcaro@iaps.inaf.it, roberto.viotti@iaps.inaf.it \\
\and Associazione Romana Astrofili, Roma, Italy, m.calabresi@mclink.it }


\abstract{ 
{\it Context:} GR\,290 (M\,33/V532=Romano's Star) is a suspected post-LBV (Luminous Blue Variable) star located in M\,33 galaxy that shows a rare Wolf--Rayet (WR) spectrum during its minimum light phase. 
In spite of many studies, its atmospheric structure, its circumstellar environment and its place in the general context of massive stars evolution is poorly known.\\ 
{\it Aims:}  Detailed study of its wind and mass loss, and study of the circumstellar environment associated to the star. \\
{\it Methods:} Long-slit spectra of GR\,290 were obtained  during its present minimum luminosity phase with the Gran Telescopio Canario covering the $\sim3600-7500$ \AA\AA\ wavelength range together with  contemporaneous BVRI photometry. The data were compared with non-local thermodynamical equilibrium (non-LTE) model atmosphere synthetic spectra computed with CMFGEN and with CLOUDY models for ionized interstellar medium regions. \\
{\it Results:} The current  $m_V=18.8$ mag, is the faintest at which this source has ever been observed.  The non-LTE models indicate effective temperature $T_{eff}=27000-30000$ K 
at radius $R_{2/3}=27-21$ ${\rm R_\odot}$ and mass loss rate $\dot{M}$=1.5$\times$10$^{-5}$ ${\rm M_\odot yr^{-1}}$. The terminal wind speed $\vv_{\infty}$=620 ${\rm km~s^{-1}}$ is faster than ever before recorded while the current luminosity $L_*=(3.1-3.7)\times$10$^5$ ${\rm L_\odot}$ is the lowest ever deduced. It is overabundant in He and N and underabundant in C and O.  It is surrounded by an unresolved compact \ion{H}{II} region with dimensions $\leq$4 pc, from where H-Balmer, \ion{He}{i} lines and $[\ion{O}{III}]$ and $[\ion{N}{II}]$ are detected. 
In addition, we find emission from a more extended interstellar medium (ISM) region which appears to be asymmetric, with a larger extent to the East ($16-40$ pc) than to the West. \\
{\it Conclusions:}  In the present long lasting visual minimum, GR\,290 is in a lower bolometric luminosity state with higher mass loss rate. The nearby nebular emission seems
to suggest that the star has undergone significant mass loss over the past $10^4-10^5$ years and is nearing the end stages of its evolution.
}


\maketitle

\section{Introduction}

  Luminous Blue variables (LBVs) are relatively short evolutionary stage in life of massive stars. 
  During this phase they lose significant mass through strong stellar winds and occasional giant eruptive events. As result, they shed their outer layers and eventually become hydrogen-deficient Wolf--Rayet (WR) stars \citep{Langer1994, HumphreysDavidson, Ekstrom, Groh2013GRB}.

   LBVs show both short timescale stochastic variability with amplitude up to few tenths of mag \citep{HumphreysDavidson,PashaLBV} 
   and long-term variability, so-called S\,Dor cycle \citep{HumphreysDavidson,Vink2012review,Humphreys2016}. 
   Typically S\,Dor cycles last for several years and their amplitude is $1-3$ mags. During S\,Dor cycle photometric variability is always accompanied 
   by changes in effective temperature, i.e. their spectral type changes in the range from B supergiants or late Of/WN stars  to  A-supergiants. 
   At the same time, the bolometric luminosity remains approximately constant \citep{Wolf1989,HumphreysDavidson}  
   or may even decrease during the visual maximum \citep{Groh2009AGCar}. 
   Moreover, Galactic objects $\eta$\,Car and P\,Cyg  have shown another kind of outbursts, so-called giant eruptions -- huge increase of bolometric 
   luminosity simultaneously with an increase of visual brightness \citep{HumphreysDavidson}.

   LBVs are divided into two subclasses: {\it low luminosity} LBVs and {\it classical} LBVs \citep{Humphreys2016}.
   The former ones originate from stars with $M_*\approx25-40\,\Msun$  reaching LBV stage after being red supergiants 
   (RSG) \citep{Humphreys2016,Meynet2005eolutionWR}. \citet{GrohMeynet2013} demonstrated that low luminosity LBVs 
   could be progenitors of Supernovae (SN) IIb. Thus low luminosity LBVs are final stage of stellar evolution, 
   rather than a transitory evolutionary phase between RSG and WR as was suggested earlier \citep{Meynet2005eolutionWR}.

   Classical LBVs originate from more massive stars ($M_*\approx40-60\,\Msun$) moving off the Main Sequence (MS) and evolving towards the WR stage via blue supergiant (BSG) and LBV phases \citep{Meynet2005eolutionWR,Meynet2011}. 
   According to \citet{Groh2014evolution}, stars with $60\,\Msun$ are spending $2\cdot10^{5}$\,years in LBVs phase, i.e., only 5\% of their lifetime. 
   Consideration of the initial mass function (IMF) leads to conclude that LBV stars are extremely rare and there should be no more than a few dozens such objects in the Galaxy. 
   Clues to  the mechanisms by which transition from O-type stars and WR stars occurs may be gleaned from a handful of objects that have been observed to display both WR and LBV  characteristics. Examples of these rare objects include  AG~Car  in our Galaxy \citep{Groh2009AGCar}, R\,127 and HDE\,269582 in Large Magellanic Cloud (\citet{Walborn2017} 
   and references therein), HD\,5980 in Small Magellanic Cloud \citep{KoenigsbergerMorrell2014} and GR\,290 in M\,33 (\citet{Polcaro2016}, and references therein).

\vskip0.2cm

   GR\,290 (M\,33-V532=Romano's Star) is located in a relatively empty field in the outskirts of M\,33 galaxy.  
   It was first classified  as a Hubble-Sandage variable by \citet{romano} and then became an
   LBV candidate \citep{HumphreysDavidson,szeifert} after introduction of this class of objects by Peter~\citet{Conti}.  Spectral variations from 
   a B-supergiant type spectrum to that of a WR of the nitrogen sequence (WN)\footnote{There is only one spectrum of GR\,290 obtained  in 1992 during  a 
   major eruption of 1992-1994 and published by  \citet{szeifert}. This spectrum may be classified as 
   that of a late type supergiant. We may assume that 
   GR\,290 has displayed a WN type spectrum since 1994, as all known spectra acquired after that time, 
   starting with the one published by \citet{Sholukhova97}, are of WN type.}, photometric  variability 
   \citep{Kurtev2001} as well as a large bolometric luminosity  \citep{Polcaro2003} are the reasons 
   that led to the promotion of GR\,290 from a LBV candidate to {\it bona fide} LBV (see \citet{Kurtev2001,Polcaro2003} and \citet{fabrika}).
   Later, \citet{polcaro10} concluded that  GR\,290's bolometric luminosity has significantly changed 
   during the minimum phase of the 2008 year and suggested that Romano's star has already passed through the LBV phase, i.e. that it is a post-LBV. 
   \citet{Humphreys2014} noted that the star does not exhibit S\,Dor transitions to the cool, dense wind state -- instead, it varies between
   two hot states on the Hertzsprung--Russell diagram, characterized by WN spectroscopic features  \citep{Humphreys2014}. 
   \citet{Humphreys2014} also, like \citet{polcaro10} earlier, suggested that Romano's star is likely in a post-LBV state. 

   The photometric investigation initiated by \citet{romano} was followed by that of \citet{Kurtev2001}, 
   who reported the photometry obtained over 1982-1990. They concluded that the eruption time-scales are 
   $\sim$20 years and that shorter timescale oscillations of $\sim$270 days or $\sim$320 days with amplitude 
   of 0.5\,mag are also present. \citet{Zharova2011} investigated photometric variability of GR\,290 using 
   both the Moscow collection of photographic plates, their own data, and the data published earlier 
   \citep{romano, HumphreysSandage, Sholukhova2002, Viotti2006}, thus constructing  the most comprehensive 
   light curve covering half a century. This light curve shows that GR\,290 exhibits irregular brightness 
   variations with different amplitudes and time scales, and where the large amplitude and complex wave-like 
   variations on the scale of several years are the most prominent.  Polcaro et al. (2016) were able to 
   further extend this light curve to the beginning of 20th century and demonstrated that GR\,290 did not 
   display any significant eruptions between 1901 and the 1960s.

   Studies of GR\,290 devoted to spectral variability show that its spectral type changes between WN11 and WN8  
   \citep{maryeva2010,polcaro10,Sholukhova2011}. Since the beginning of 2000s it has made this transition twice \citep{Polcaro2016}.
   \citet{Viotti2006,Viotti2007} first described an anticorrelation between emission line strength and brightness. 
   Then \citet{maryeva2010} found a correlation of spectral changes and the visual brightness typical for LBVs: the brighter it is, 
   the cooler the spectral type. However, among all known LBVs only HD\,5980 convincingly shows a hotter spectrum in the minimum of brightness.

   The models of GR\,290 atmosphere have been constructed for three states -- luminosity maximum of 2005 
   and the minimum of brightness in 2008 \citep{maryeva2012}, and at the moderate luminosity maximum of 2010 \citep{ClarkLBV2012}. 
   \citet{Polcaro2016} built nine models for the most representative spectra acquired between 2002 and 2014 and demonstrated how the 
   GR\,290 wind structure varies correlated with brightness changes -- the slow and dense wind at brightness maxima becomes faster 
   and thinner at minima. Similar correlations appear also in HD\,5980 \citep{Georgiev2011} and provide clues to the mechanisms that 
   drive the spectral transitions. However, due to its faintness in recent years, it has been challenging to obtain the spectra  
   of GR\,290 with high enough quality for wind speeds to be adequately estimated.
   
   Many LBVs are surrounded by small nebulae whose structure and chemical composition have been studied in order to investigate their past history
   (e.g. \citet{MartayanLobel2016} and references therein). In this regard, \citet{fabrika} claimed a marginal detection, in an H$\beta$ image,
   of a compact elongated nebula around GR\,290. But no detailed spectrophotometric study of the nebular emission near GR\,290 has been so far carried out. 
   
\begin{figure*}
{\centering \resizebox*{2\columnwidth}{!}{\includegraphics[angle=0]{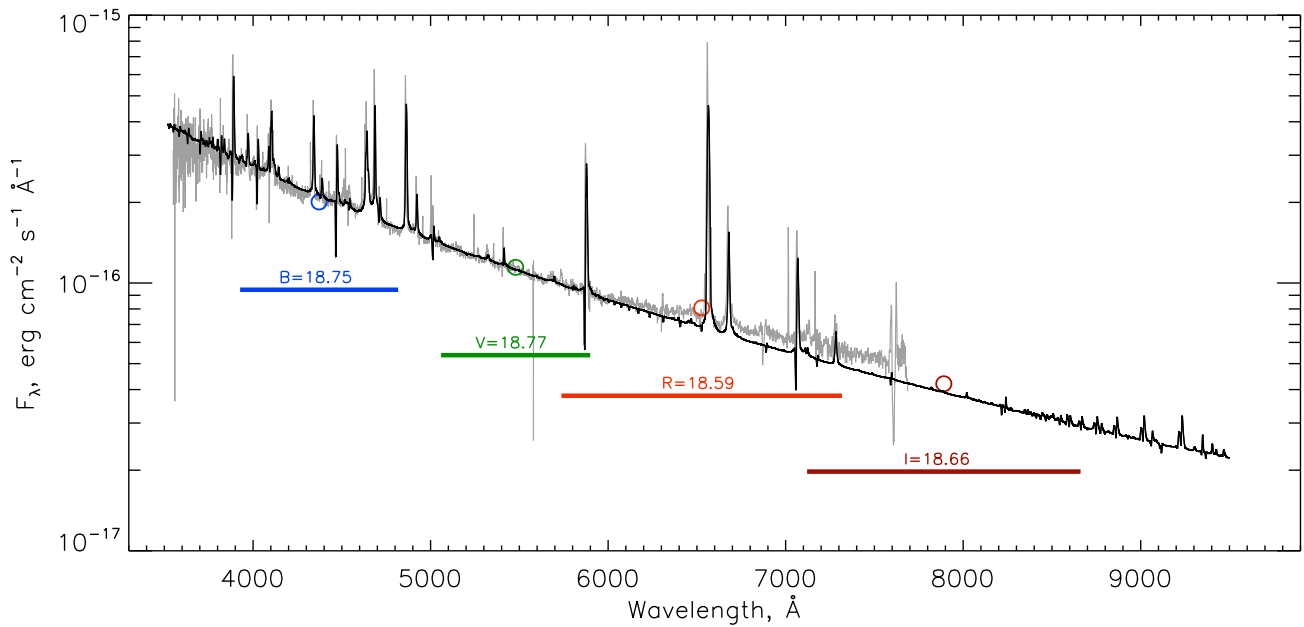}}}
\caption{Flux-calibrated spectrum of GR\,290 (grey line) compared with the best-fit $\beta$=2 CMFGEN
model (black line). Measured photometric magnitudes and corresponding bandwidths are indicated. 
The model spectrum is scaled for the distance to the M\,33 
and the interstellar extinction ($E_{(B-V)}=0.06$) is applied to it. The adopted distance to M\,33 is 847$\pm$61 kpc (distance module 
24.64 $\pm$0.15) from \citet{Galleti2004}.  
\label{fig:model_sed}
}
\end{figure*}

   In this paper we present the results of our ongoing spectral and photometric monitoring of GR\,290. 
   We report the results of observations performed in 2016 with one of the best spectral resolutions and signal-to-noise ratio ever obtained for 
   this object. We are particularly targeting the information on the circumstellar emission around GR\,290 that may be extracted from these spectra, 
   which may give us some knowledge on the mass loss during previous stages of the star's life.

   In section 2 we describe the observations; in section 3 the current spectral characteristics and
   deduced parameters from CMFGEN fits; in section 4 the spectral characteristics of the 
   interstellar medium and OB-star regions in the neighborhood of GR\,290. In section 5 we 
   present the conclusions.

\section{Observations}

GR\,290 was observed with the OSIRIS spectrograph on the {\it Gran Telescopio Canarias (GTC)}
on 2016 July 30 and August 27 (MJD 57599.1569, 57627.073, respectively) in Service Observing mode 
using the R2500-set of gratings and the long slit (7.4') with a slit 
width of 0.6''. These gratings cover the ultraviolet ($U$), visual ($V$) and red ($R$) wavelength
intervals as defined in the third column of Table~\ref{table_obs}. This table also lists  
the filter ID, the exposure time, the range in signal-to-noise (S/N), the spectral dispersion and the 
resolution for each of the observations. Position angles were $94.75\deg$ and $95.2\deg$,  respectively. 

\begin{table}
\begin{center}
\begin{footnotesize}
\begin{small}
\caption{Spectroscopic observations 2016 July and August. \label{table_obs}}
\begin{tabular}{lccrlcl}
\hline
\hline
\\
{\bf MJD} & {\bf Filter ID} & {\bf $\lambda$-range } &{\bf $t_{exp}$} &{\bf S/N} &{\bf \AA/pix} &{\bf Res.}    \\
          &                 &       {(\AA)}          &   (s)          &          &              &            \\
\hline
\hline
\\
57599.16     &R2500U   &    3440 - 4610   &       1200       &   40-50    & 0.62  & 2555   \\
57627.07     &  "      &       "          &       1200       &  20-40     &     " &  "         \\
57599.17     &R2500V   &     4500 - 6000  &        900       &   25-46    & 0.80  & 2515        \\
57627.09     &  "      &       "          &        900       &  25-50     &     " &  "        \\
57599.18     &R2500R   &    5575 - 7685   &        600       &   30-37    & 1.04  &  2475        \\
57627.10     &  "      &       "          &        600       &  20-27     &     " &  "        \\
\hline
\hline
\end{tabular}
\end{small}
\end{footnotesize}
\end{center}
\end{table}

    The spectra were pre-processed by the standard GTC/OSIRIS pipeline\footnote{http://gtc-osiris.blogspot.mx/2012/10/the-osiris-offline-pipeline-software.html}
    (Ederoclite \& Cepa 2010). The data were then reduced 
    using the standard IRAF procedures~\footnote{IRAF is distributed by the NOAO, which is operated by AURA, 
    under contract with NSF.}, which included the bias and flat field corrections as well as the wavelength 
    calibration. The rms of the fit for the wavelength calibration on the comparison lamps is
    0.032\,\AA\ in the red and 0.015\,\AA\ in visual and blue spectra. Lines lying on overlapping segments
    obtained from the 3 filters were checked for relative shifts. These lines are \ion{He}{i} 4471\,\AA\ which
    appears in both the $U$ and the $V$ segments and the \ion{He}{i} 5876\,\AA\ which appears in the $V$ and $R$
    segments. Both lines have P\,Cyg profiles and we measured the emission and the absorption components 
    separately as individual features, for the purpose of verifying the wavelength calibration. We find that 
    the $U$ and $V$ segments overlap to within $\pm$0.2\,\AA\ at $\lambda$4471 in both the July and August 
    observations. The $V$ and $R$ segments overlap to within $\pm$0.3\,\AA\ in the August observation. However, 
    \ion{He}{i} 5876\,\AA\ line in the $R$ segment in the July observation is shifted by about $-1$\,\AA\, the source of 
    which we are unable to ascertain. Although such a significant shift is not present in other lines of this
    R segment, we deem the RVs from this segment to be less reliable than those in the others.

    The flux calibration was performed using standard routines in IRAF, as well as IDL procedures originally 
    written for SCORPIO\footnote{ SCORPIO is Spectral Camera with Optical Reducer for Photometric and
    Interferometric Observations (SCORPIO) \citep{AfanasievMoiseev2005} on the Russian 6-m telescope} data reduction.
    The resulting flux-calibrated spectra, however, still suffer from the following limitations: 1) the lack of 
    an accurate and updated extinction curve of the site, mainly in the red and 2) the fact that the standard star that 
    was used is more than 10 magnitudes brighter than GR\,290 so that the instrument was de-focused in order 
    to avoid saturation. The absolute fluxes (shown in Fig.~\ref{fig:model_sed}) are thus to be 
    considered only approximate.

\begin{table}
\begin{center}
\begin{small}
\caption{Photometry obtained on 2016 Jul 31 using the Johnson-Cousins filters.  \label{table_phot}}
\begin{tabular}{lllrl}
\hline
\hline
\\
{\bf Band} & {\bf magnitude  }   & {\bf $\pm$ } &{\bf S/N}  &{\bf UT}     \\
\hline
\hline
\\
 V      &18.77   &      0.04 &    71 &   01:44:14      \\
 R      &18.59   &      0.04 &    72 &   01:53:10      \\
 I      &18.66   &      0.05 &    46 &   02:00:57      \\
 B      &18.75   &      0.03 &    99 &   02:10:43      \\
\hline
\hline
\end{tabular}
\end{small}
\end{center}
\end{table}

    Photometric observations in the Johnson-Cousins filters were obtained using the 1.52-m Cassini Telescope 
    run by INAF-Osservatorio Astronomico di Bologna in Loiano 
    on 2016 July 31, providing a photometric reference for the GTC July 30 observation. Table~\ref{table_phot} lists 
    in column 1 the wavelength band, in column 2 the magnitude value, in column 3 and 4 the measurement 
    uncertainty and the S/N, respectively, and in column 5 the Universal Time (UT) of the observations.  
    With $m_V=18.8\,{\rm mag}$, GR\,290 is currently at its faintest since the beginning of the 20th 
    century (see \citet{Polcaro2016} for the photometric history).


\begin{table*}
\begin{center}
\begin{small}
\caption{Emission line radial velocity, FWHM and equivalent width measurements. \label{table_RVs_final} }
\begin{tabular}{llllllllll}
\hline
\hline
\\
 &   &\multicolumn{2}{c}{\bf ---- July ----}  &\multicolumn{2}{c}{\bf ---- August ---} &\multicolumn{4}{c}{\bf ----------- Average ------------- }\\
\multicolumn{2}{c}{ID}  &{RV}&{FWHM} & {RV} &{FWHM} &{RV$_{em}$}& {-EW}& { V$_{abs}$}&$V_{edge}$ \\
\hline
\hline
\\
\ion{He}{I} & 3704.00?&         -15. &      332. &      -56. &      307.& 135    &  1.1    &  -210 & -250      \\
\ion{H}{I}  & 3750.20 &        -129. &      328. &      ...  &      ... & 16     &  0.6    &  -172 & -283:     \\
\ion{H}{I}  & 3770.60 &        -102. &      358. &      -85. &      374.&  75    &  1.8    &  -258 & -400:     \\
\ion{H}{I}  & 3797.90 &        -105. &      308. &      ...  &      ... & -10    &  0.4    &  -221 & -310      \\
\ion{He}{I} & 3819.60*&        -178. &      361. &     -136. &      228.&  17    &  1.2    &  -252 & -530      \\
\ion{H}{I}  & 3835.40*&        -147. &      320. &     -104. &      461.&  32    &  1.2    &  -201 & -370:     \\
\ion{He}{I} & 3888.65 &        -149  &      457  &     -183  &      485:&  20    & 10.4    &  -350 & -660      \\
\ion{H}{I}  & 3970.10*&        -124. &      325. &     -149. &      310.&  18    &  2.2    &  -193 &  ...      \\
\ion{He}{I} & 4026.20*&        -149. &      342. &     -152. &      342.&  10    &  2.4    &  -248 & -550      \\
\ion{N}{IV} & 4057.76 &        -223. &      273. &     -164. &      148.& -10    &  0.5     &  ... &  ...      \\
\ion{Si}{IV} &4088.70*&        -182. &      213. &     -173. &      198.&   2    &  1.7    &   ... &  ...      \\
\ion{N}{III} &4097.30 &        -200. &      154. &     -190. &      ... & -4     &  2.0:   &  -230 & -438      \\
\ion{H}{I}   &4101.74 &         -67. &      373. &      -51. &      285.& 121:   &  5.2    &  ...  &  ...      \\
\ion{Si}{IV} &4116.10*&        -228. &      364. &     -211. &      233.& -32    &  2.8    &  ...  &  ...      \\
\ion{N}{III} &4195.76 &        -234. &      141. &     -266. &      164.& -66:   &  0.2    &  ...  &  ...      \\
\ion{N}{III} &4200.10 &        -162. &      206. &     -165. &      142.&  13    &  0.6    &  -159:& -260:     \\
\ion{H}{I}  & 4340.50*&        -149. &      380. &     -119. &      366.&  42    &  7.9    &  ...  &  ...      \\
\ion{N}{III} &4379.20*&        -186. &      260. &     -164. &      226.&  -7    &  1.0    &  ...  &  ...      \\
\ion{He}{I}  &4387.90*&        -198. &      301. &     -135. &      246.&  13    &  1.6    &  ...  &  ...      \\
\ion{He}{I}  &4471.50*&        -149. &      389. &     -127. &      362.&  28    &  4.9    &  -275 & -630      \\
\ion{N}{III} &4534.60*&        -194. &      178. &     -149. &      165.& -22    &  0.4:   &  -233:& -280:     \\
\ion{He}{II} &4541.60 &        -145. &      145. &      ...  &      ... &  29:   &  2.0:   &  -182 & -300:     \\
\ion{N}{III} &4634.14 &        -183  &      256  &     -188  &      252 & -9     &  5.5    &  ...  &  ...      \\
\ion{N}{III} &4640.64 &        -197. &      252. &     -195. &      258.& -16    &  9.4    &  ...  &  ...      \\
\ion{He}{II} &4685.6 *&        -176. &      333. &     -174. &      326.&   7    & 17.4    &  ...  &  ...      \\
\ion{He}{I}  &4713.1 *&        -146. &      197. &     -130. &      407.&  28    &  1.6:   &  -291 & -500:     \\
\ion{H}{I}  & 4861.3 *&        -171. &      395. &     -167. &      401.&  11    & 18.2    &  ...  &  ...      \\
\ion{He}{I} & 4921.9 *&        -185. &      335. &     -174. &      371.&   4    &  4.3    &  ...  &  ...      \\
\ion{He}{I} & 5015.7 *&        -162. &      275. &     -149. &      287.&  15    &  1.7    &  -200:&  ...      \\
\ion{He}{I} & 5047.7 *&        -174. &      184. &     -143. &      237.&  41    &  0.6    &  ...  &  ...      \\
\ion{He}{II} &5411.52 &        -150. &      260. &     -169. &      294.&  15    &  1.9    &  -217 & -433      \\
\ion{He}{I}  &5875.66*&        -175. &      444. &     -171. &      464.&   5    & 24.5    &  -304 & -655      \\
\ion{H}{I}  & 6562.80 &        -184. &      301. &     -176. &      269.&  -1    & ...     &  ...  &  ...      \\ 
\ion{H}{I}  & 6562.80*&        -181. &      466. &     -172. &      461.&   8    & 68.0    &  ...  &  ...      \\ 
\ion{He}{I} & 6678.15*&        -176. &      417. &     -185. &      395.&  -6    & 18.0    &  ...  &  ...      \\
\ion{He}{I}  &7065.20*&        -152. &      467. &     -157. &      450.&  13    & 18.1    &  -306 & -560      \\
\ion{N}{IV}  &7109.35 &        -169. &      181. &      ...  &      ... &   8    &  0.2:   &  ...  &  ...      \\
\ion{N}{IV} & 7122.98 &        -160. &      231. &      ...  &      ... &  ...   &  0.3    &  ...  &  ...      \\
\hline
\hline
\end{tabular}
\end{small}
\end{center}
{\bf Notes:} {\small Columns: 1 lists the primary atomic identification and corresponding laboratory
wavelength in \AA; 2 and 4 list the RVs for, respective, the July and the August
observations after applying the heliocentric corrections of $+27~{\rm km~s^{-1}}$ for July and
$+24~{\rm km~s^{-1}}$ for of August; 3 and 5 are the corresponding full-widths at half maximum intensity.
Columns 6-9 list measurements performed on the combined July+August spectrum
after correcting for the adopted -179 ${\rm km~s^{-1}}$ systemic velocity of M\,33. Column 6 is the RV of the
emission lines; column 7 is the corresponding equivalent width; column 8 is the centroid of the P\,Cyg
absorption component when present; and column 9 is the velocity of the P\,Cyg absorption edge (i.e., the
location where the absorption reaches the continuum). RVs and FWHM's are in units of ${\rm km~s^{-1}}$; EWs are
in units of \AA, and the negative sign in the column heading indicates emission. Asterisk indicates
the line was used to determine the RV of the system.}
\end{table*}
\begin{figure}
\includegraphics[width=0.45\linewidth]{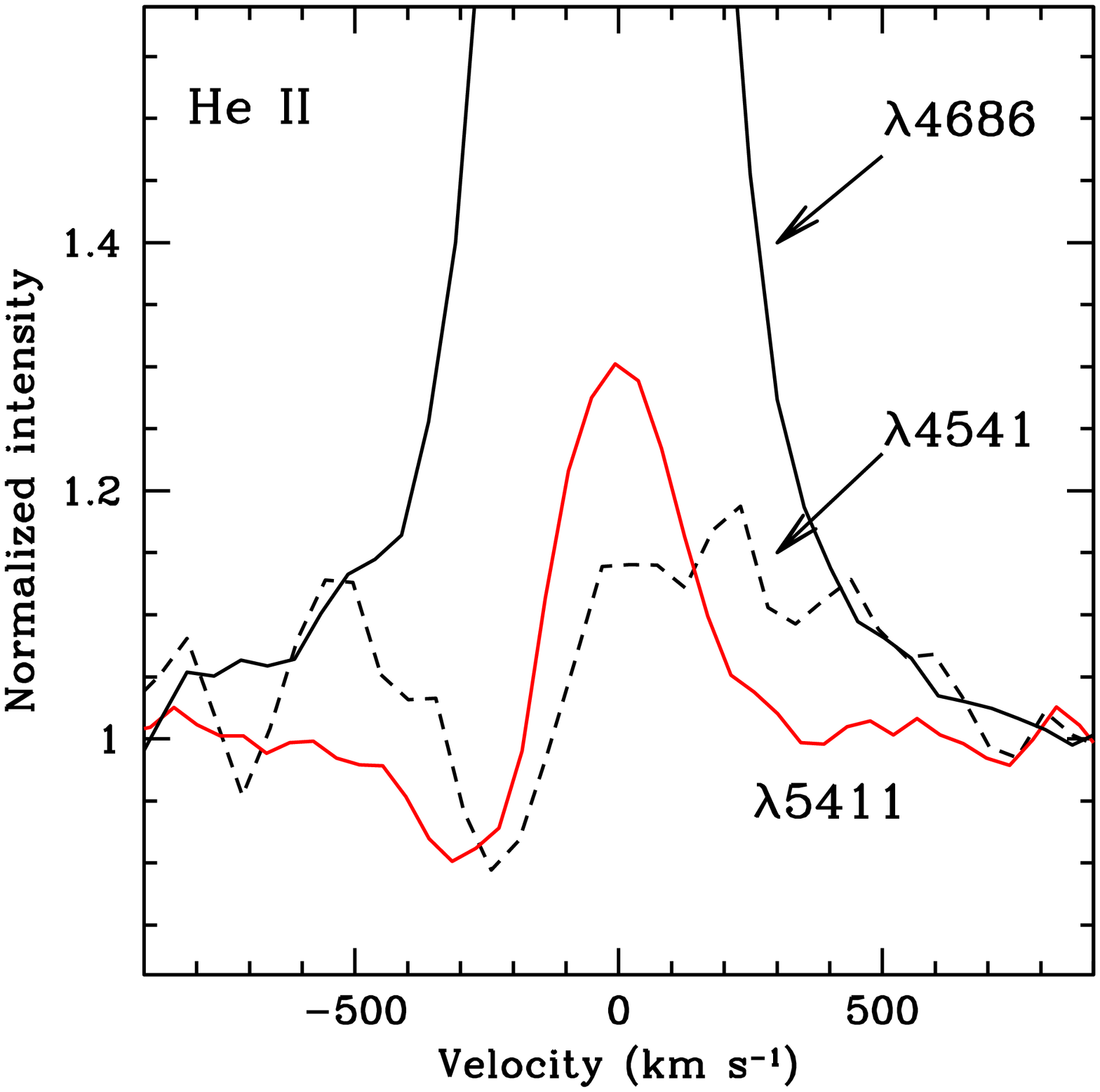}
\includegraphics[width=0.45\linewidth]{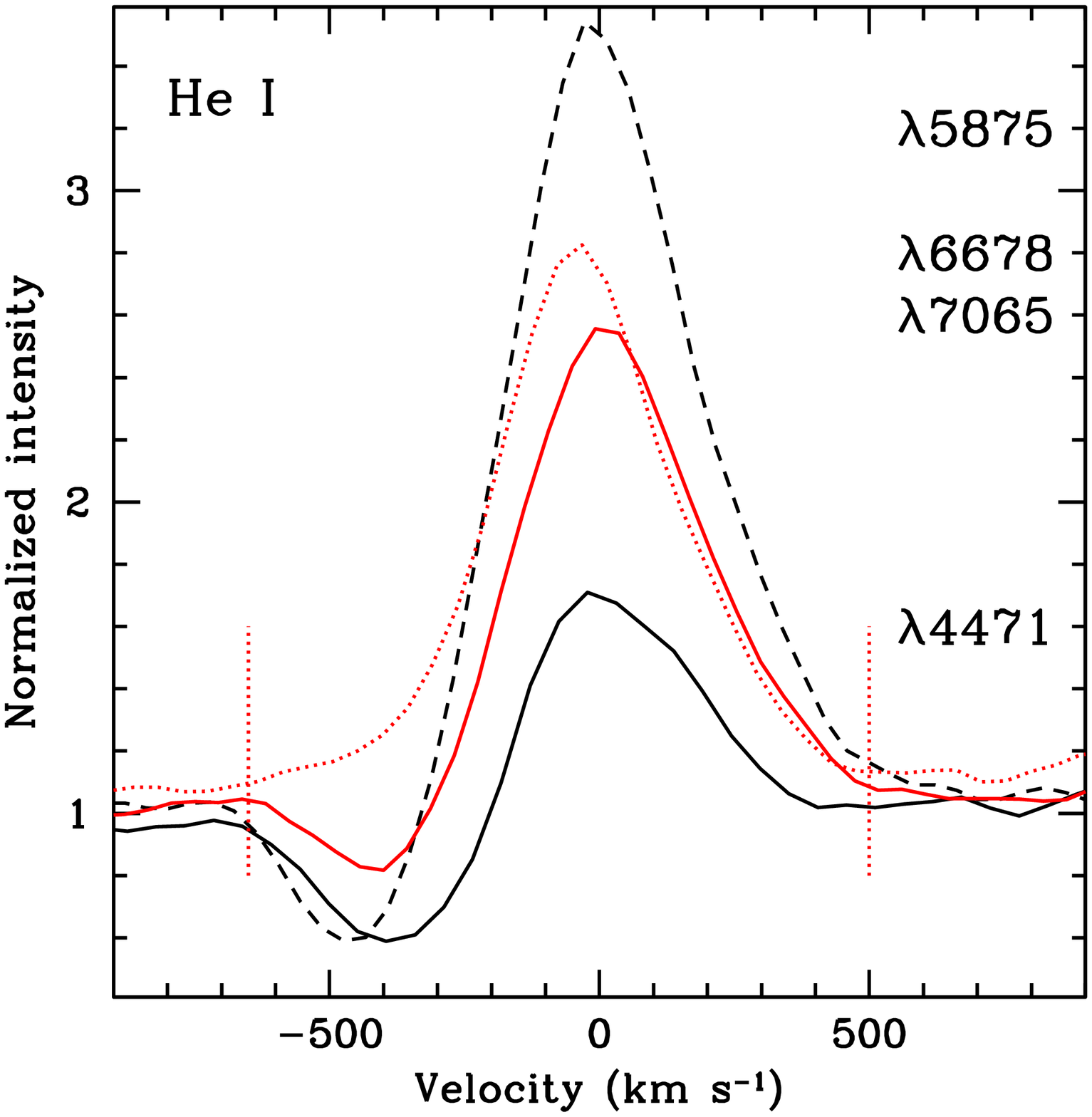}
\caption{{\it Left:} Line profiles of \ion{He}{ii} plotted on a velocity scale corrected 
for the adopted -179 ${\rm km~s^{-1}}$ systemic velocity. The continuous line shows the strong $\lambda$4686, with its
extended electron scattering wings and $\lambda$5411  with its strong P\,Cyg absorption. The
dash line is $\lambda$4541  which is contaminated on the blue wing with \ion{N}{III}.  {\it Right:}
\ion{He}{i} lines profiles of $\lambda$4471 (continuous, black), $\lambda$5876 (dash, black), $\lambda$6678 (dot, red)
and $\lambda$7065 (continuous, red). The vertical lines indicate the approximate locations where the line meets the 
continuum level. \label{fig_HeII_HeI}
}
\end{figure}
\begin{figure}
\includegraphics[width=0.48\linewidth]{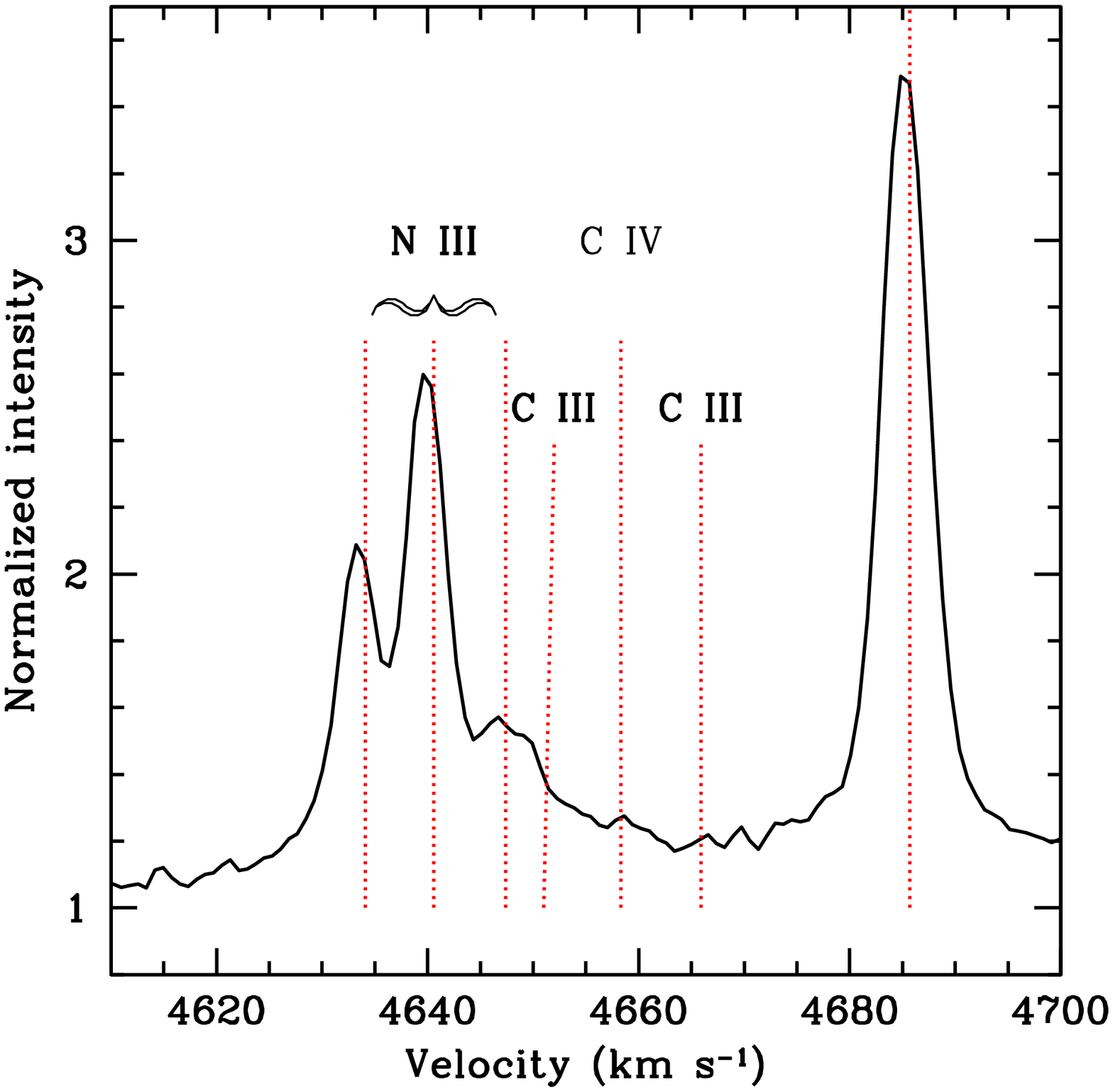}
\includegraphics[width=0.48\linewidth]{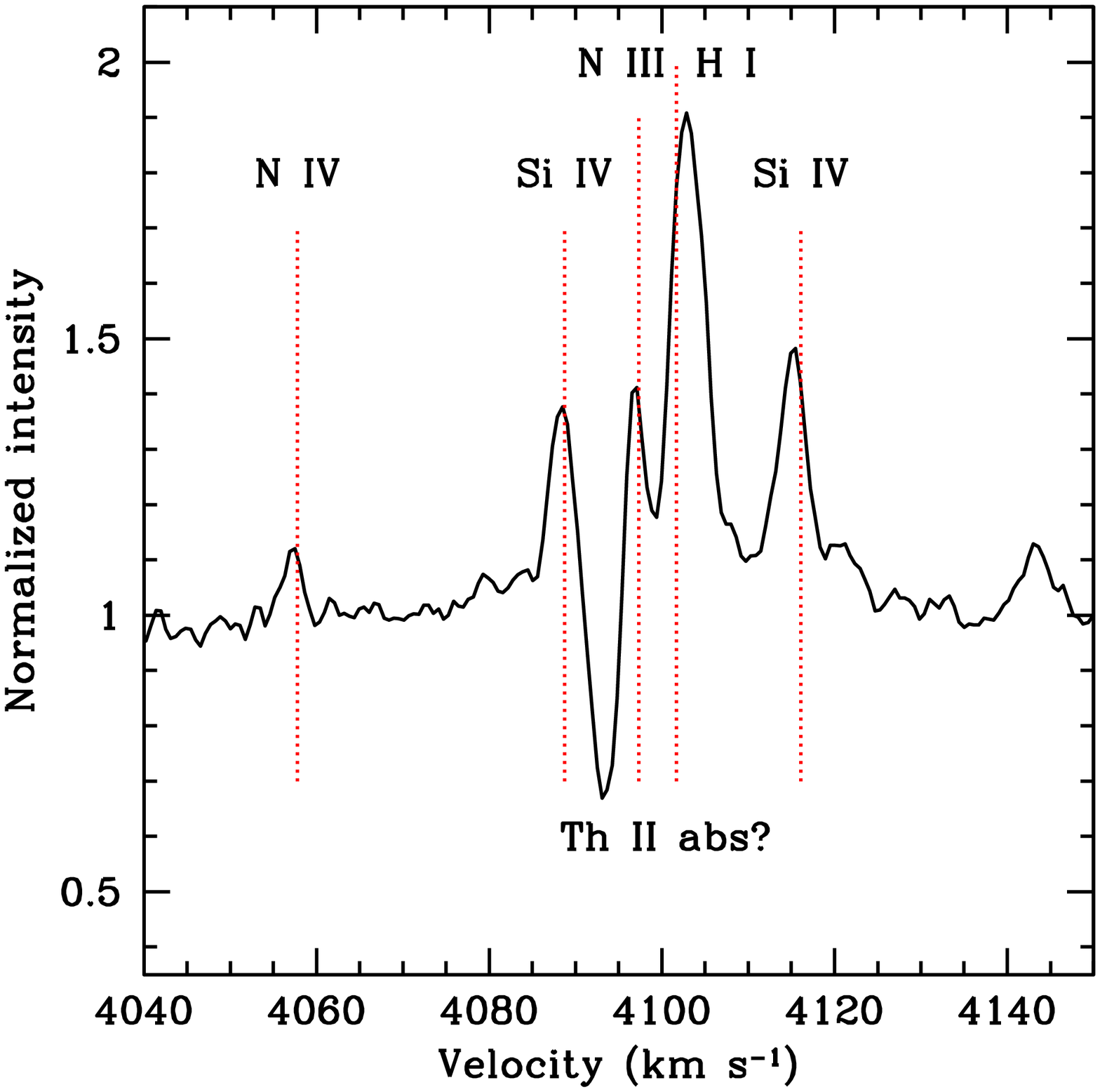}
\caption{\label{fig: niii_4686} Line profiles on a wavelength scale corrected for the adopted
-179 ${\rm km~s^{-1}}$ systemic velocity of M\,33. {\it Left:} \ion{N}{III} 4634-41 and \ion{He}{ii} 4686 emission lines. Note
the possible contribution of \ion{C}{IV}. {\it Right:} \ion{N}{IV} 4058 and the 4100 \AA\ blend that includes \ion{Si}{IV}, \ion{N}{III}
and \ion{H}{I} lines. Vertical lines indicate the laboratory wavelength of the transitions.
}
\end{figure}

\section{Spectrum of GR\,290}

  The flux calibrated spectrum of GR\,290 is characterized by a blueward rising continuum  
  and strong emission lines, as illustrated in Fig.~\ref{fig:model_sed}. In addition to H$\alpha$ 
  and H$\beta$, the spectrum is dominated by lines from \ion{He}{i}, \ion{He}{ii} and \ion{N}{III}. 
  The P\,Cyg absorption components of the \ion{He}{i} triplet lines are enhanced, with respect to the singlets, due to
  the overpopulation with respect to LTE of the low-lying triplet levels in the low density stellar wind. 
  The \ion{He}{II} Paschen-like line at 4686\,\AA\ (3-4)   is observed in emission, as in earlier spectra, 
  while the multiplet 2 Brackett-like lines at 5411\,\AA\ (4-7) and 4541\,\AA\ (4-9) have P\,Cyg profiles 
  and are now stronger than in earlier spectra. The \ion{He}{ii} line from this same series at 4199\,\AA\ (4-11) 
  (blended with \ion{N}{III}) also displays a P\,Cyg profile. The profiles of some of the \ion{He}{ii} 
  lines are illustrated  in Fig.~\ref{fig_HeII_HeI}.    

  The strongest \ion{N}{III} lines are those at 4634-41\,\AA, which are partially resolved
  (Fig.~\ref{fig: niii_4686}, left). There is a prominent absorption blueward of \ion{N}{III} 4097\,\AA\ 
  suggesting that this line also has a P\,Cyg profile. The \ion{N}{III} 4196, 4200 and  4379\,\AA\ 
  are also now present, consistent with a higherd egree of ionization in the wind suggested by the increased 
  strength of \ion{He}{ii} 5411.  

  Most importantly, we note the presence for the first time of \ion{N}{IV} 4058 \AA\ in Fig.~\ref{fig: niii_4686} (right).  
  It is clearly present in both the July and the August spectra and has an equivalent width of 0.5 \AA\ on the
  combined spectrum. The \ion{N}{IV} lines at 7109 and 7122 \AA\ appear to also be weakly present. 

  The primary indicator of ionization classification given by \citet{SmithSharaMoffat1996} is the ratio of 
  equivalent widths (EW) of \ion{He}{II} 5411/\ion{He}{I} 5876\,\AA. We find that this ratio is $\leq$0.1,
  implying a degree of ionization corresponding to WN8. The earlier criteria of \citet{vanderHucht1981},
  which are based on the \citet{Smith1968} system, rely on the ratios of \ion{N}{III}/\ion{N}{IV}/\ion{N}{V} 
  4634-4641, 4057 and 4604-4620\,\AA, respectively. The fact that EW(\ion{N}{III}) $\gg$ EW(\ion{N}{IV}) 
  is consistent with WN8 or cooler (see Table~\ref{table_RVs_final}). 

\begin{figure}
\includegraphics[width=\linewidth]{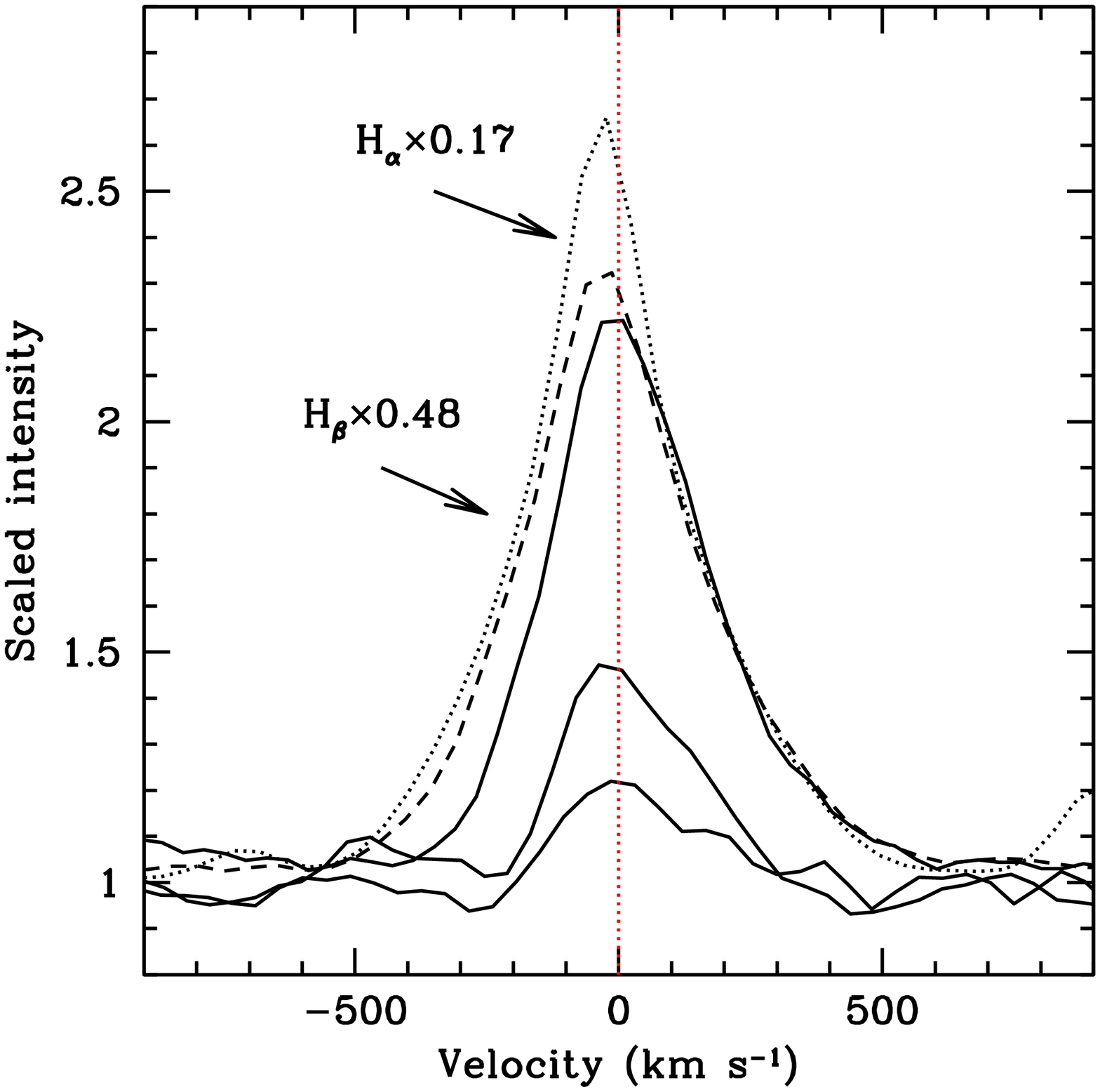}
\caption{Line profiles of the Balmer-H sequence plotted on a velocity scale corrected for the adopted 
        -179 ${\rm km~s^{-1}}$ systemic velocity. The lines of H$\alpha$ and H$\beta$ are scaled by 
        factors of 0.17 and 0.48, respectively, so as to align the red wings with that of H$\gamma$.  
        The continuous lines show H$\gamma$, H$\epsilon$ and \ion{H}{I} 3835\,\AA. The progressive red-shift at 
        shorter wavelengths appears to be a consequence of the increasing strength of the blue-shifted  
        absorption which, in the helium lines, takes on the form of P\,Cyg absorption. This effect is 
        explained by the fact that the optical depth of the wind is highest for H$\alpha$ and H$\beta$ 
        and decreases for lines having smaller transition probabilities.}
        \label{fig_montage_H}
\end{figure}
\subsection{Radial Velocities}

   The radial velocity (RV) and full-width at half maximum intensity (FWHM) was obtained through
   a Gaussian fit to individual line profiles or, in the case of overlapping lines and P\,Cyg profiles, 
   through simultaneous multiple Gaussian fits.
   Table \ref{table_RVs_final} contains the results of our measurements. Columns 1 and 2 list the primary 
   atomic identification and corresponding laboratory wavelength, Columns 3 and 5 list the heliocentric RV
   of the line centroid for, respective, the July and the August observations. The corresponding heliocentric
   corrections are $+27~{\rm km~s^{-1}}$ for July and $+24~{\rm km~s^{-1}}$ for August. Columns 4 and 6 list the 
   corresponding values of the FWHM. 

   In order to obtain the velocity of GR\,290 with respect to M\,33, we averaged the RVs of the lines in 
   Table \ref{table_RVs_final} marked with an asterisk. The selection of these lines is based on the following 
   criteria: 1) relatively unblended line; 2) relatively good S/N; 3) little or no ambiguity in the identification.  
   We find average RV(July)=-171$\pm$22 (s.d.) ${\rm km~s^{-1}}$ and RV(August)=-155$\pm$24 (s.d.) ${\rm km~s^{-1}}$. 
   Averaging the RVs of July and August yields RV(GR\,290)=-163$\pm$32 ${\rm km~s^{-1}}$.
   The three averages are consistent, within the uncertainties, with each other and with the heliocentric  
   velocity -179$\pm$3 ${\rm km~s^{-1}}$ of M\,33 \footnote{NASA/IPAC/JPL extragalactic Database}. We thus 
   adopt this latter value for the systemic velocity of GR\,290 in the remainder of this paper.  
   The standard deviations are of similar value to our estimated measurement uncertainty of $\pm$30 ${\rm km~s^{-1}}$.
   
   Since we do not detect variability in our two spectra that is larger than the uncertainty discussed above,
   we combined the July and August spectra, shifted the average by +179 ${\rm km~s^{-1}}$ and 
   remeasured the lines. We list in Table \ref{table_RVs_final} the RV of the emission in Column 7, the equivalent
   width in Column 8, the RV of the absorption component in the case of P\,Cyg profiles in Column 9, and the
   intersection of the P\,Cyg absorption with the continuum level in Column 10. The measurements for P\,Cyg profiles
   were performed with two-Gaussian fits.\footnote{It is important to note that two-Gaussian fits to P\,Cyg profiles
   lead to a significant difference in the centroid and FWHM compared to a procedure in which the emission 
   and absorption components are each measured independently. This is because when deblending, the combined 
   profile is fit with an absorption and a neighboring, partially overlapping emission. This results in an 
   emission component having a bluer RV and a broader FWHM than if it were fit individually.} We note that 
   the contribution of nebular emission affects the FWHM. For example, in the case of $H\alpha$, we measure 
   FWHM=6.6 \AA\ if the line is fit with a single Gaussian while FWHM=10.2 \AA\ if the line is treated as a 
   blend of a broad stellar line and a narrow superposed nebular emission. Despite this large difference in FWHM, 
   the RV in both cases is nearly the same.
   
   The average $V_{edge}$=-400$\pm$144 ${\rm km~s^{-1}}$, is similar to the value of $\vv_\infty$ listed in
   Polcaro et al. (2016) for recents spectra. However, the \ion{He}{I} lines have $V_{edge}\simeq$620 ${\rm km~s^{-1}}$, 
   implying wind speeds that are faster than ever before observed.
   
   The Balmer-H lines also show the signs of a relatively fast outflow, as can be seen from the growing 
   asymmetry in the blue wing of the higher excitation lines of this series. This is illustrated in 
   Fig.~\ref{fig_montage_H} where H$\alpha$ and H$\beta$ were scaled in intensity so as to align the red
   wings with that of H$\gamma$. The other lines plotted are H$\epsilon$ and \ion{H}{i} 3835\,\AA. This asymmetry results 
   from the same process that leads to P\,Cyg absorptions in \ion{He}{i} but because of the large optical depth of the 
   H-lines, a fully developed absorption component is not visible.

\begin{figure*}
{\centering \resizebox*{0.62\columnwidth}{!}{\includegraphics[angle=0,viewport=45 12 554 536,clip]{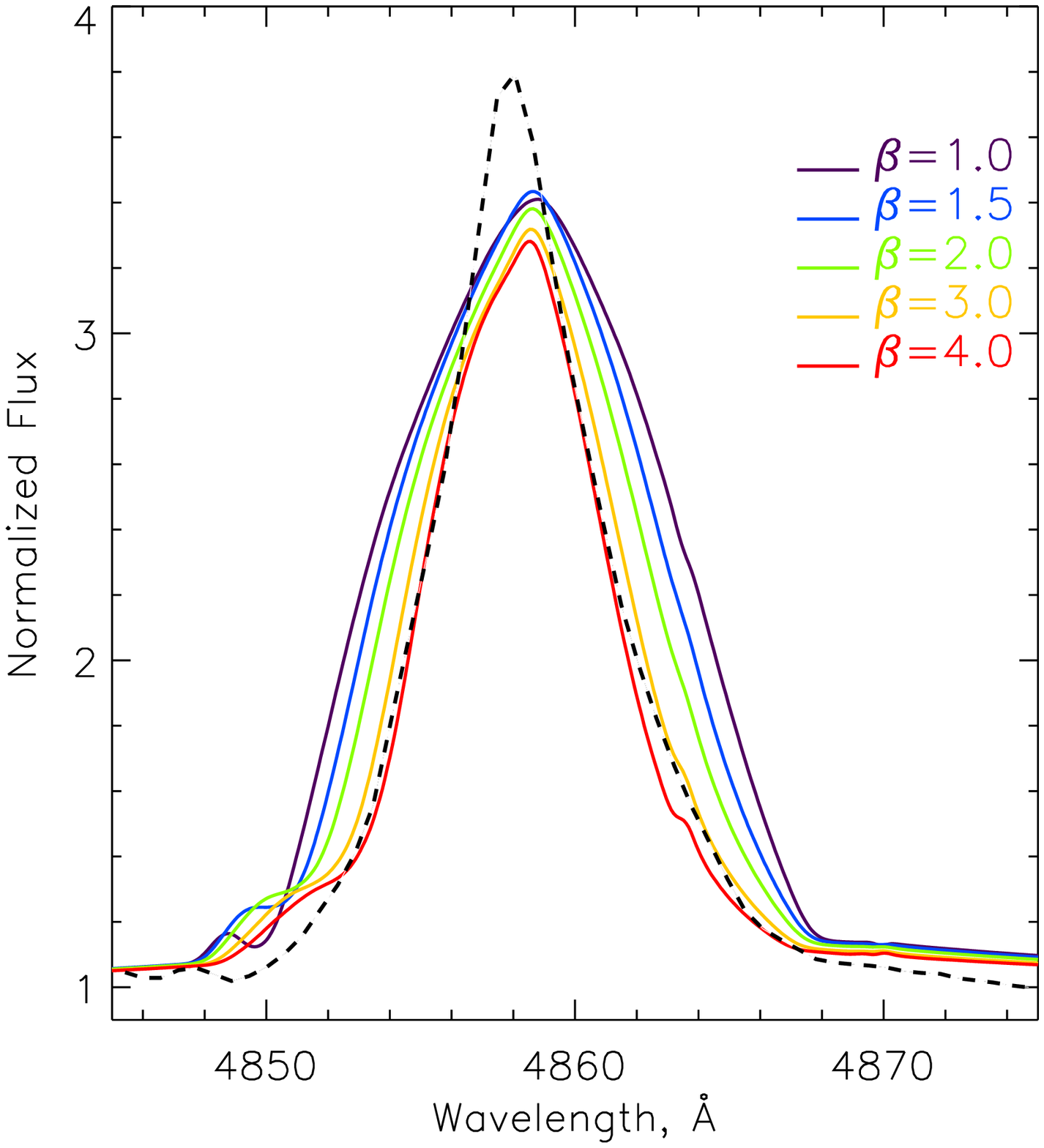}}}
{\centering \resizebox*{0.62\columnwidth}{!}{\includegraphics[angle=0,viewport=45 12 554 536,clip]{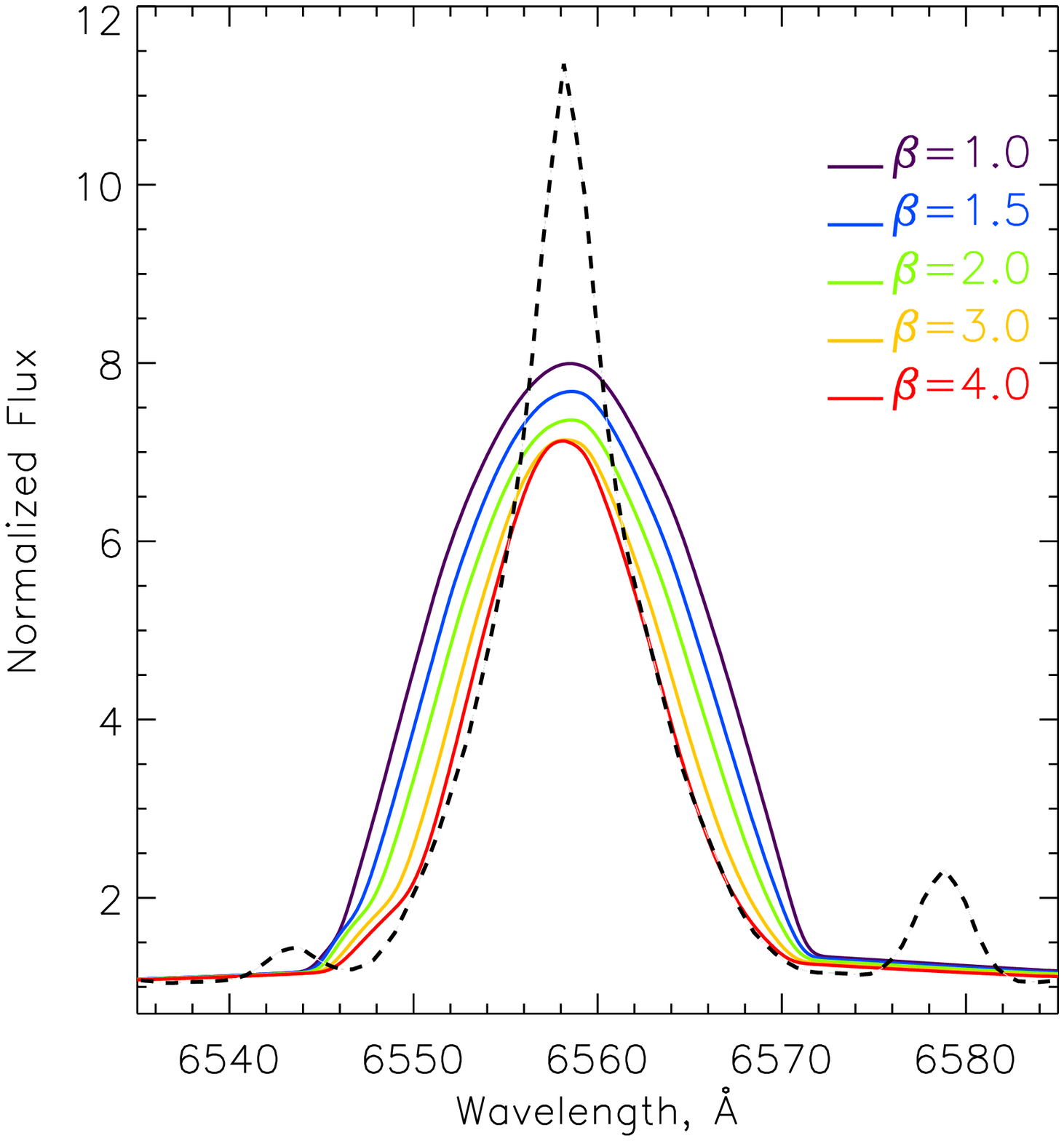}}}
{\centering \resizebox*{0.62\columnwidth}{!}{\includegraphics[angle=0,viewport=45 12 554 536,clip]{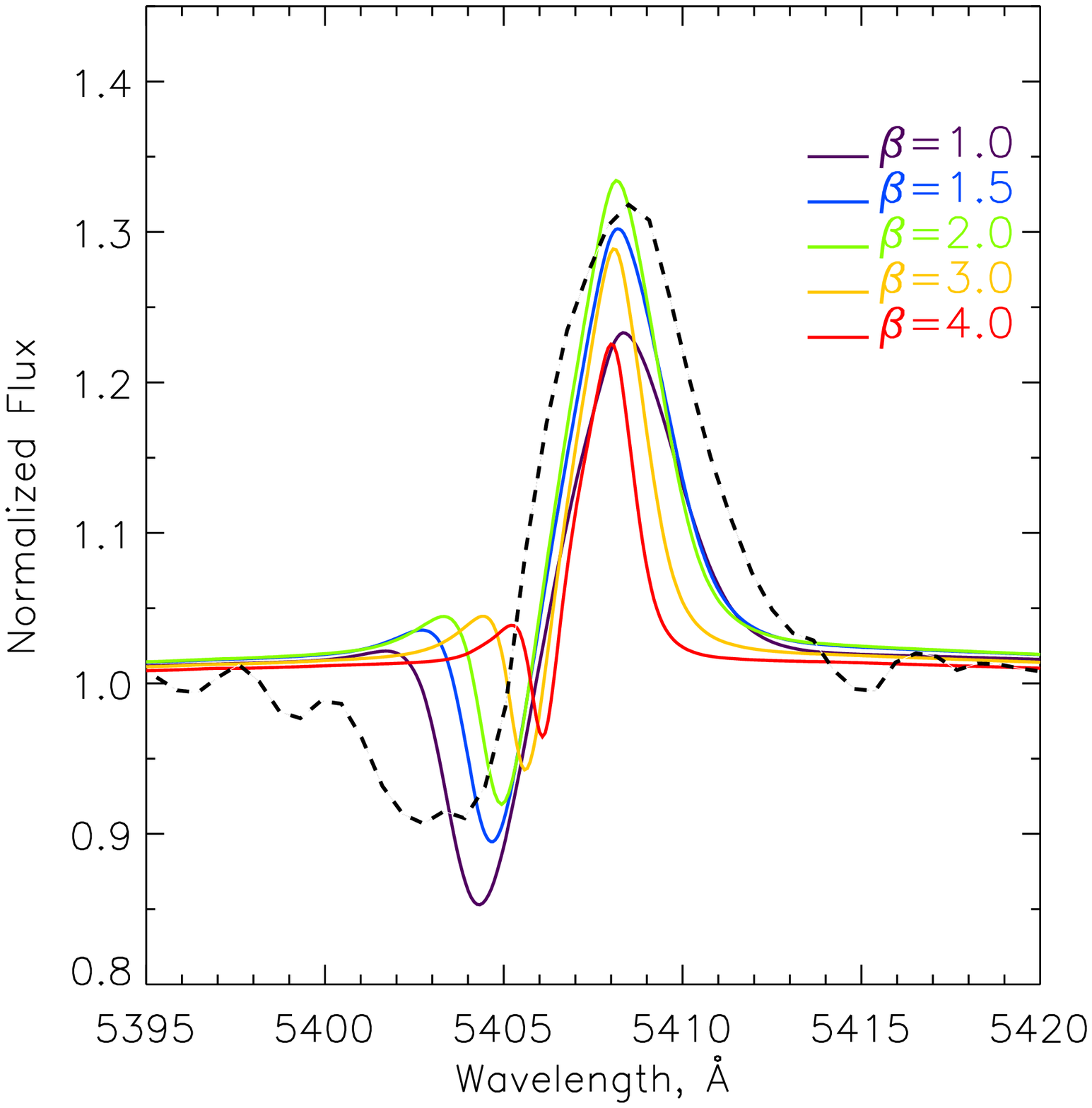}}}
\caption{ The effect of increasing $\beta$-parameter in the velocity law on spectral line profiles. 
From left to right -- modeled profiles of H$\beta$, H$\alpha$ and \ion{He}{II} 5411. For comparison, the actual profiles 
observed in the spectrum of GR\,290 are shown by black dashed line.}
\label{fig:beta}
\end{figure*}
\begin{figure*}
{\centering \resizebox*{2\columnwidth}{!}{\includegraphics[angle=0,viewport=85 18 1240 380,clip]{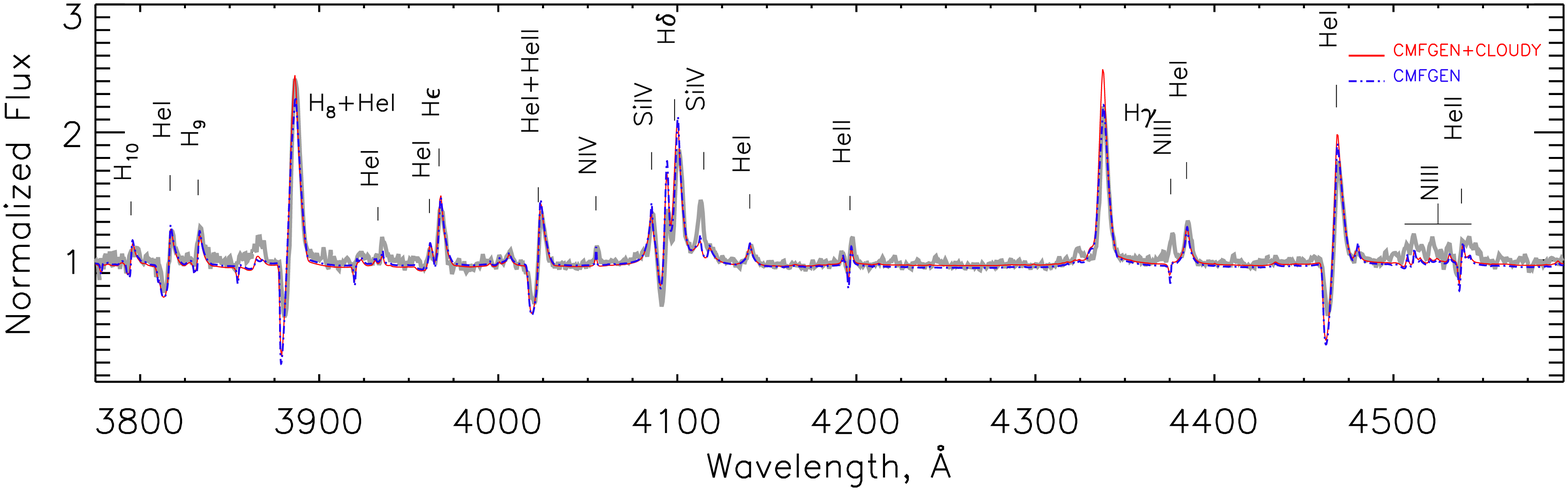}}}
{\centering \resizebox*{2\columnwidth}{!}{\includegraphics[angle=0,viewport=85 18 1240 380,clip]{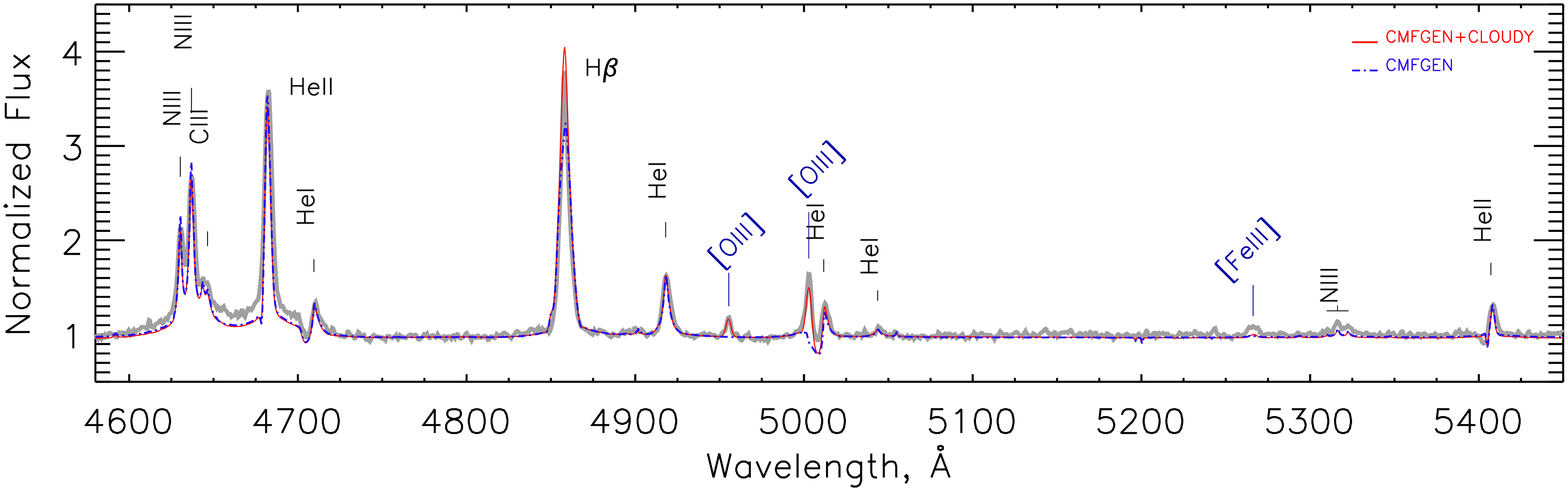}}}
{\centering \resizebox*{2\columnwidth}{!}{\includegraphics[angle=0,viewport=85 18 1240 380,clip]{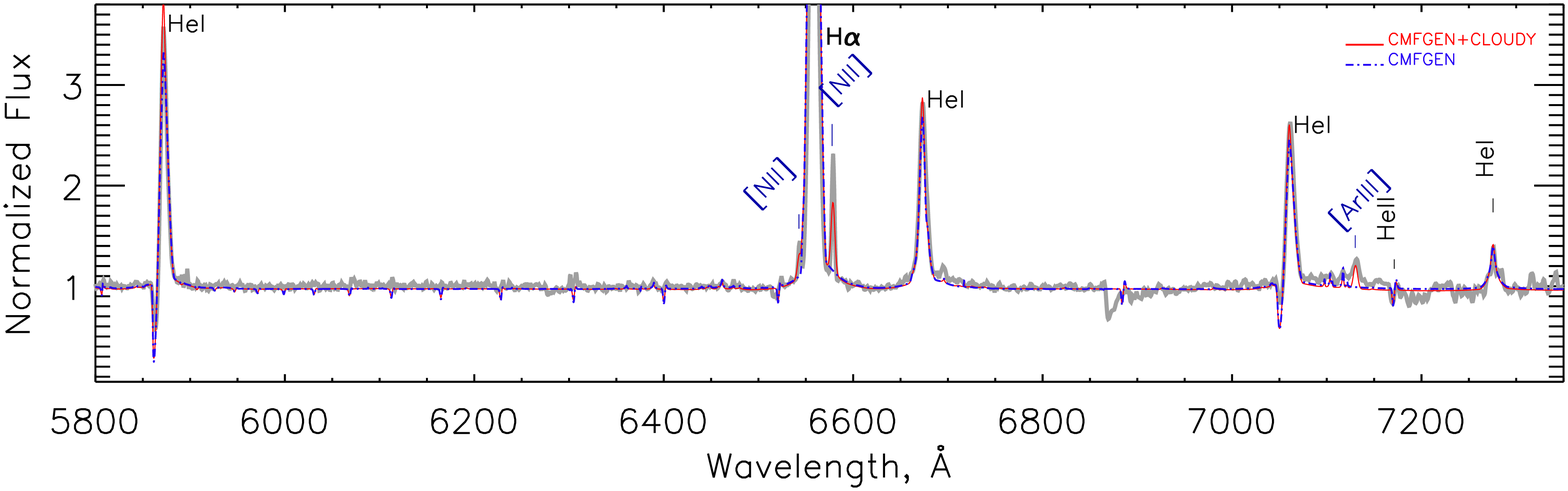}}}
\caption{Normalized optical spectrum of GR\,290 (grey line) compared with the best-fit CMFGEN model with 
$\beta=2$ (blue dash-dotted line). Red solid line shows the sum of CMFGEN (stellar atmosphere) and CLOUDY (surrounding nebula) models. 
The CLOUDY model has increased value of oxygen abundance (in comparison with the CMFGEN model), which we discuss in the text. 
}
\label{fig:modelsspectra}
\end{figure*}
\begin{table}
\begin{center}
\begin{small}
\caption{Chemical abundances in the CMFGEN model. \label{table_abundances}}
\begin{tabular}{lllll}
\hline
\hline
\\
{\bf Species}& {\bf Rel. Num. Frac.}             & {\bf Mass Frac }         &    NH            &{\bf $Z/Z_\odot$} \\
\hline
\hline
\\
     H       &     2.2                           &    0.35                  &   12              &   0.5       \\
     He      &     1.0                           &    0.64                  &   11.66           &   2.3       \\
     C       &     $1\times 10^{-4}$             &    $2\times10^{-4}$      &   7.66            &   0.06      \\
     N       &     $1.5\times 10^{-3}$           &    $3.5\times10^{-3}$    &   8.83            &   3.0       \\
     O       &     $1\times10^{-4}$              &    $2.5\times10^{-4}$    &   7.66            &   0.03      \\
\\
     Ne      &     $2.4\times10^{-4}$            &    $7.8\times10^{-4}$    &   8.04            &  0.45     \\
     Mg      &     $5.0\times10^{-5}$            &    $2.0\times10^{-4}$    &   7.37            &  0.3      \\
     Al      &     $8.4\times10^{-6}$            &    $3.7\times10^{-5}$    &   6.58            &  0.65     \\
     Si      &     $3.0\times10^{-5}$            &    $1.4\times10^{-4}$    &   7.14            &  0.2      \\
     S       &     $2.3\times10^{-5}$            &    $1.2\times10^{-4}$    &   7.02            &  0.32     \\
     Ar      &     $6.5\times10^{-6}$            &    $4.2\times10^{-5}$    &   6.47            &  0.4      \\
     Ca      &     $4.2\times10^{-6}$            &    $2.7\times10^{-5}$    &   6.28            &  0.4      \\
     Fe      &     $4.9\times10^{-5}$            &    $4.4\times10^{-4}$    &   7.4             &  0.32     \\
\hline
\hline
\end{tabular}
\end{small}
\end{center}
\end{table}

\subsection{Stellar wind models \label{cmfgen}}

   We used the CMFGEN atmospheric modeling code \citep{Hillier5} in order to estimate the current stellar parameters of GR\,290. 
   This code solves the radiative transfer equation for objects with spherically symmetric extended outflows using either the Sobolev 
   approximation or the full comoving-frame solution of the radiative transfer equation. CMFGEN incorporates line blanketing, 
   the effect of Auger ionization and clumping, which is parametrized by the volume filling factor $f=\bar{\rho}/\rho(r)$, 
   where $\bar{\rho}$ is the homogeneous (unclumped) wind density and ${\rho}$ is the density inside clumps assumed to be optically thin 
   to radiation, while the interclump medium is void \citep{HillierMiller1999}. Every model is defined by a hydrostatic stellar radius 
   $R_*$, luminosity $L_*$, mass-loss rate $\dot{M}$, filling factor $f$,wind terminal velocity $\vv_\infty$, stellar mass $M$ and by 
   the abundances $Z_i$ of the chemical elements that are included.    
    LBVs may  have inflated radii and pseudophotospheres \citep{Grafener2012}, which may cause large uncertainties in the the $R_*$ value derived from CMFGEN. 
    Thus,   hydrodynamical models need to be used to compute the hydrostatic radius and when discussing the evolutionary status of these objects, which goes 
    beyond the scope of our current investigation.

   The CMFGEN models were computed in an iterative manner, starting from an initial model and then adjusting parameters until a reasonable 
   match between the model and observed spectra was attained. The temperature $T_{\rm{eff}}$\footnote{$T_*$ and $T_{\rm eff}$ are effective 
   temperatures at the bottom of the wind and at radius $R_{2/3}$ where Rosseland optical depth is $2/3$.} was first adjusted by using the 
   ratios of \ion{He}{II}/\ion{He}{I} and \ion{N}{II}/\ion{N}{III}/\ion{N}{IV}. Then the luminosity was adjusted by comparing with the contemporary 
   $V$-band and $B-V$ photometry (see Table 2). The value of $\dot{M}$ was obtained from the line intensities, and the $\vv_\infty$ from 
   the P\,Cyg profile shapes. The filling factor $f$=0.15 was determined by fitting emission line profiles. 

   The abundances of chemical elements used in the models are listed in Table \ref{table_abundances}. The relative number fraction 
   is given in column 2, the corresponding mass fraction in column 3,the adopted solar abundance in column 4 and the model abundance 
   relative to the solar abundance in column 5. The abundances of H, He, C, N, and O were estimated from the model, 
   while for the abundances of Ar, S and Ne we used those of M\,33 from \citet{magri10}. For the abundances of other elements 
   (Mg, Al, Si, Ca and Fe) we assumed a values of half solar as the metallicity of the outer regions of M\,33 is low.

   Beta velocity law $\vv(r)=\vv_\infty\left(1-\frac{R_*}{r}\right)^\beta$ approximation is one of the basic simplifications 
   typically adopted while constructing atmospheric models of hot stars. Radiation-driven wind theory \citep{Puls1996,LamersCassinelliBook} 
   predicts the values of $\beta=$~0.5-1, while more hydrodynamically consistent numerical models favor $\beta=2.4$ for inner parts of the wind, 
   and $\beta=0.9$ -- for the outer ones \citep{Sander2017beta}. In our previous paper by \citet{Polcaro2016} we built the models assuming 
   $\beta=1$, as our main goal was to compare the spectra acquired with different spectral resolutions and covering different spectral ranges 
   in order to track the evolution of temperature, luminosity and mass loss rate over time. In the present article we are studying the spectrum 
   with much better resolution covering the whole visible range, and immediately after starting the modeling process we have found that the profiles 
   of hydrogen emission lines are not consistent with $\beta=1$. Therefore we had to build the models for several $\beta$ values. Fig.~\ref{fig:beta} 
   illustrates the changes of lines' profile with increasing $\beta$. 
   
\begin{table}
\begin{center}
\begin{small}
\caption{CMFGEN model fit results. \label{table_CMFGEN}}
\begin{tabular}{llll}
\hline
\hline
\\
     {\bf Quantity }                           & {\bf Model 1  }           & {\bf Model 2  }             & {\bf Note }   \\
\hline
\hline
\\
${\rm L_*/10^5\times\,L_\odot}$                 & $(3.7\pm0.2)$             &  $(3.1\pm0.2)$              &               \\
${\rm log  (L_*/L_\odot)}$                     & $5.57\pm0.03$             &  $5.49\pm0.03$              &               \\
${\rm \dot{M}/10^{-5}\times\,M_\odot~yr^{-1}}$  & $(1.5\pm0.2)$             &  $(1.4\pm0.2)$              &               \\
${\rm R_*/R_\odot}  $                          & $15-16.5$                 &  $13.5-15.2$                & (a)           \\
${\rm T_*/K}  $                                & $36 000\pm500$            &  $36 000\pm500$             & (a)           \\
${\rm R_{2/3}/R_\odot}$                        & $21-24.2$                 &  $22.6-26.6$                & (b)           \\
${\rm T_{\rm eff}/K}$                          & $30 000\pm700 $           &  $27 500\pm700 $            & (b)           \\
${\rm \vv_\infty/km\,s^{-1}}$                  & $620\pm50$                &  $620\pm50$                 &               \\
${\rm \beta } $                                & $2.0$                     &  $3.0$                      &               \\
\hline
\hline
\end{tabular}
\end{small}
\end{center}
{\bf Notes:} {\small (a)  The radius and temperature at the bottom of the wind where $\tau=18.7$;    (b) The temperature and radius at $\tau=2/3$.}
\end{table}

\begin{figure*}
{\centering \resizebox*{0.325\columnwidth}{!}{\includegraphics[angle=0,viewport=35 12 335 314,clip]{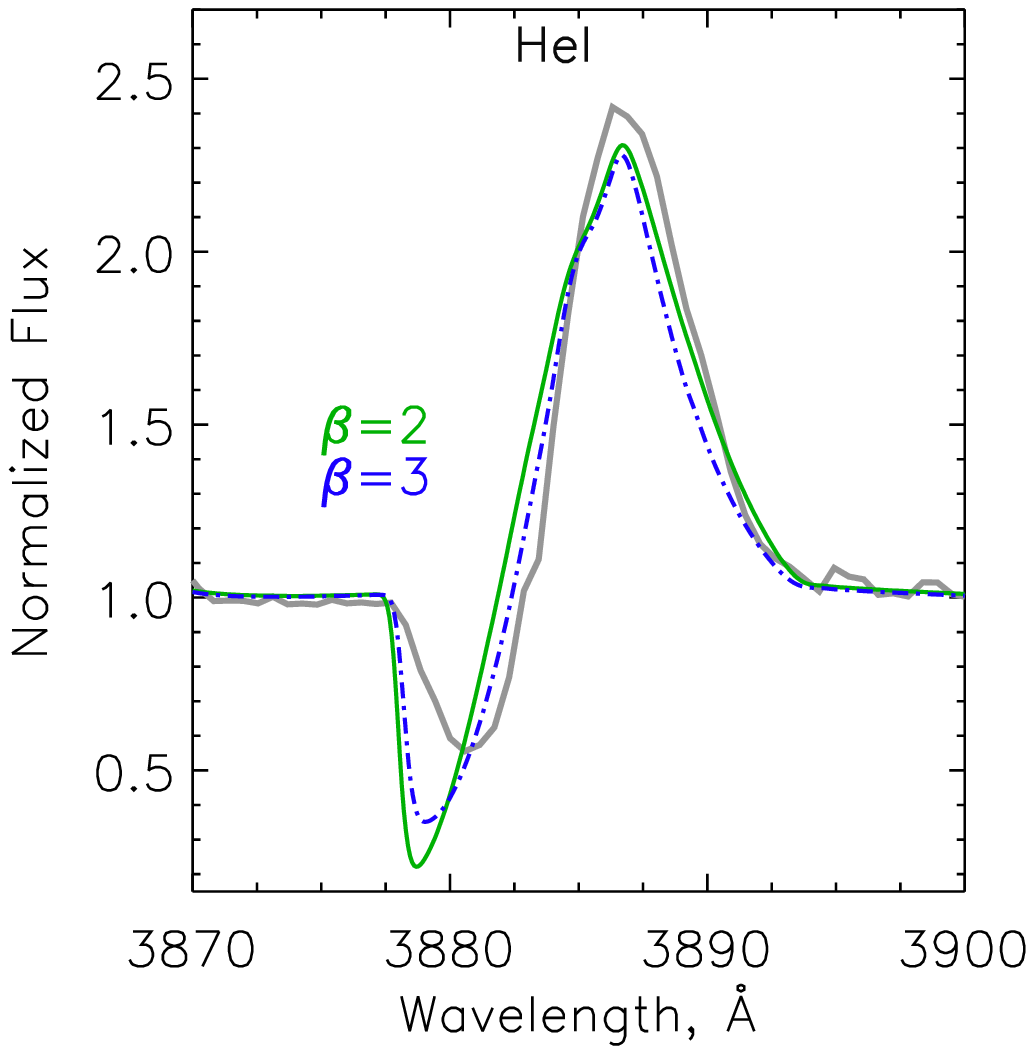}}}
{\centering \resizebox*{0.325\columnwidth}{!}{\includegraphics[angle=0,viewport=35 12 335 314,clip]{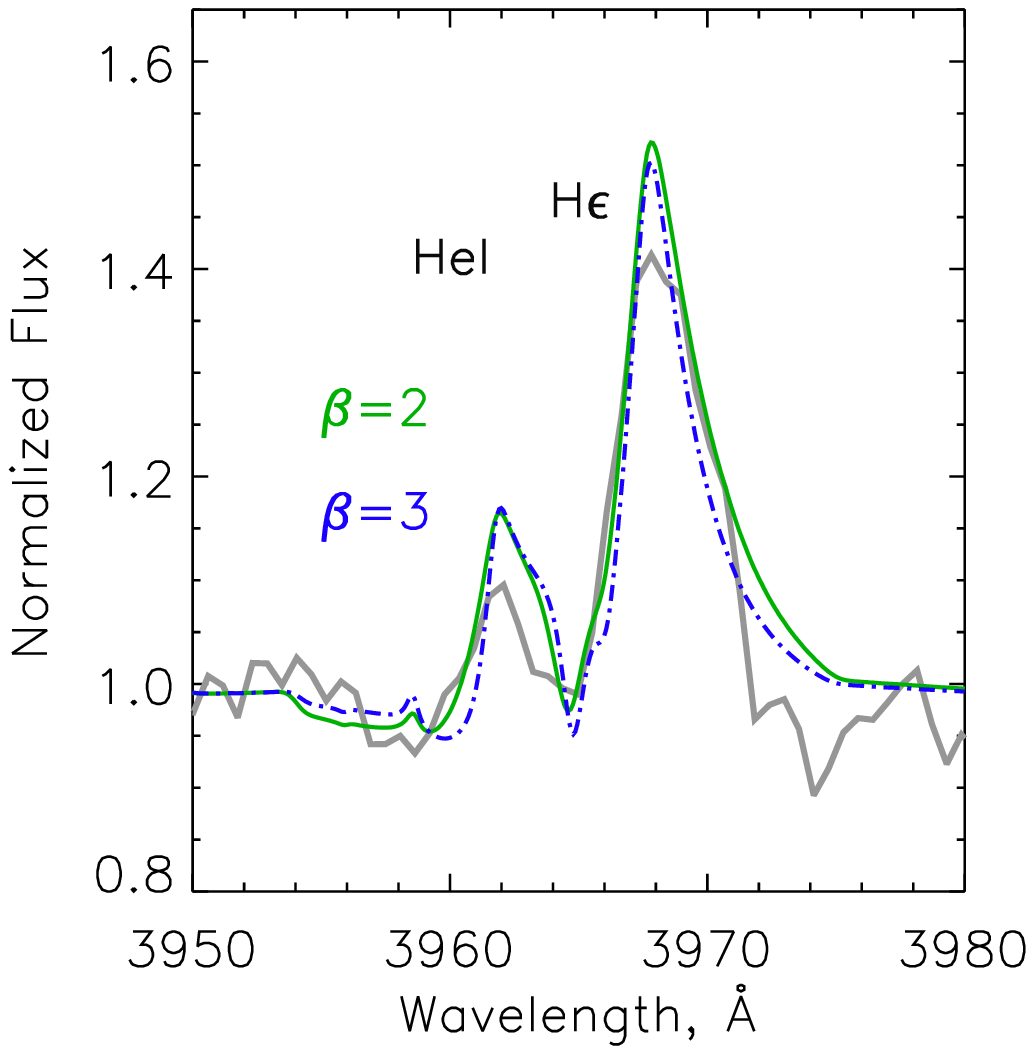}}}
{\centering \resizebox*{0.325\columnwidth}{!}{\includegraphics[angle=0,viewport=35 12 335 314,clip]{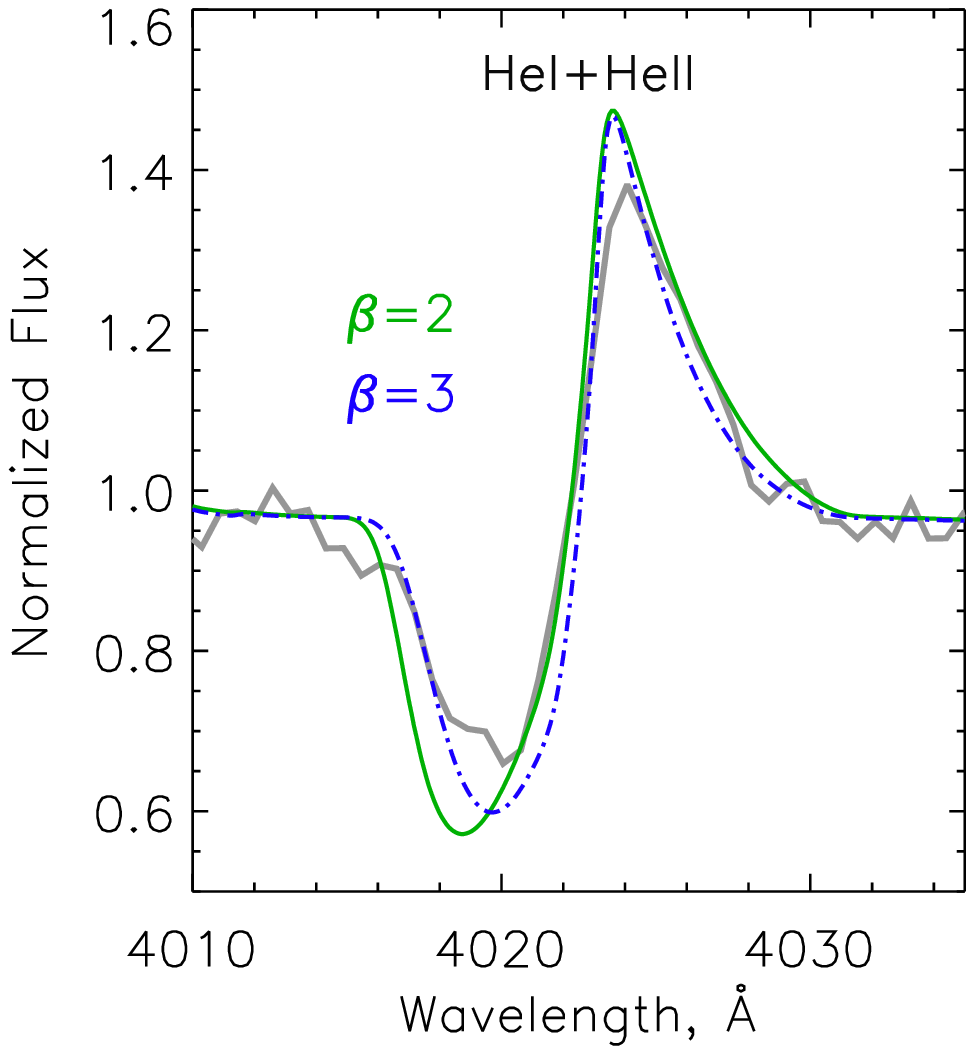}}}
{\centering \resizebox*{0.325\columnwidth}{!}{\includegraphics[angle=0,viewport=35 12 335 314,clip]{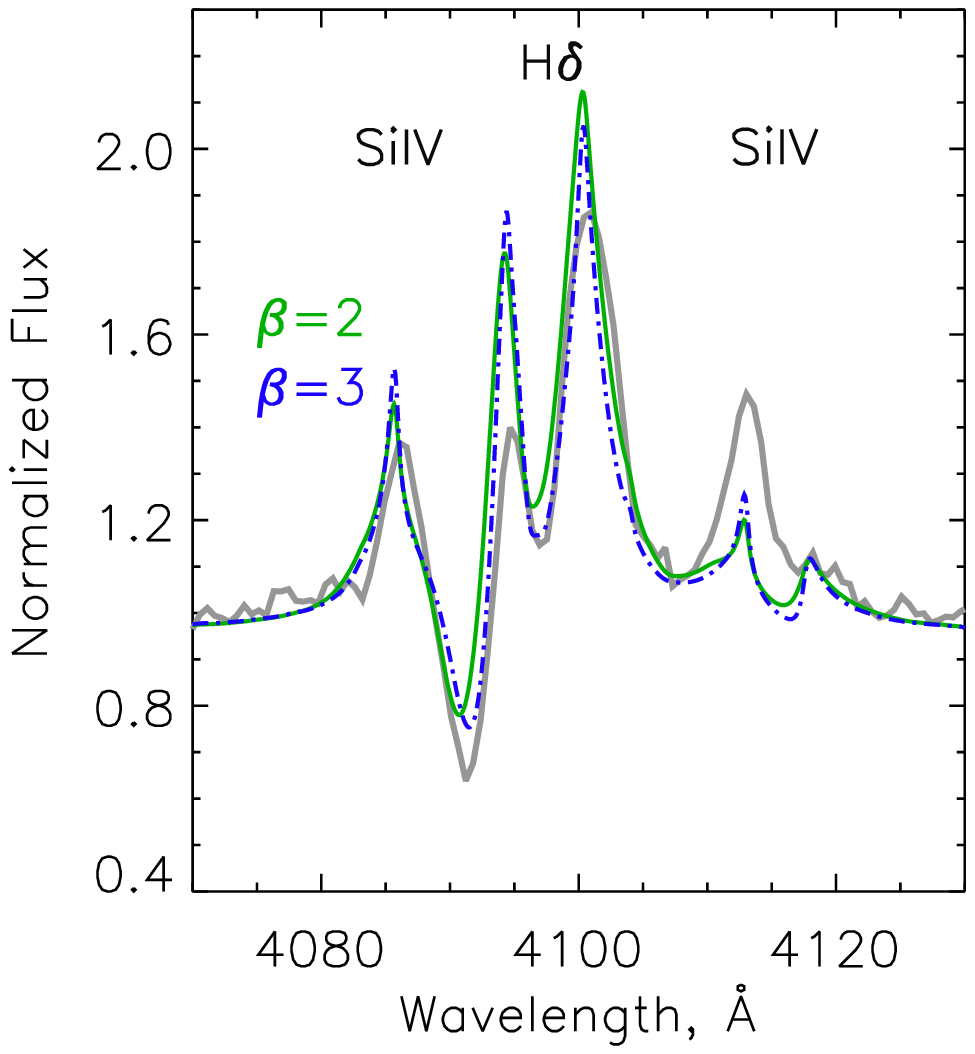}}}
{\centering \resizebox*{0.325\columnwidth}{!}{\includegraphics[angle=0,viewport=35 12 335 314,clip]{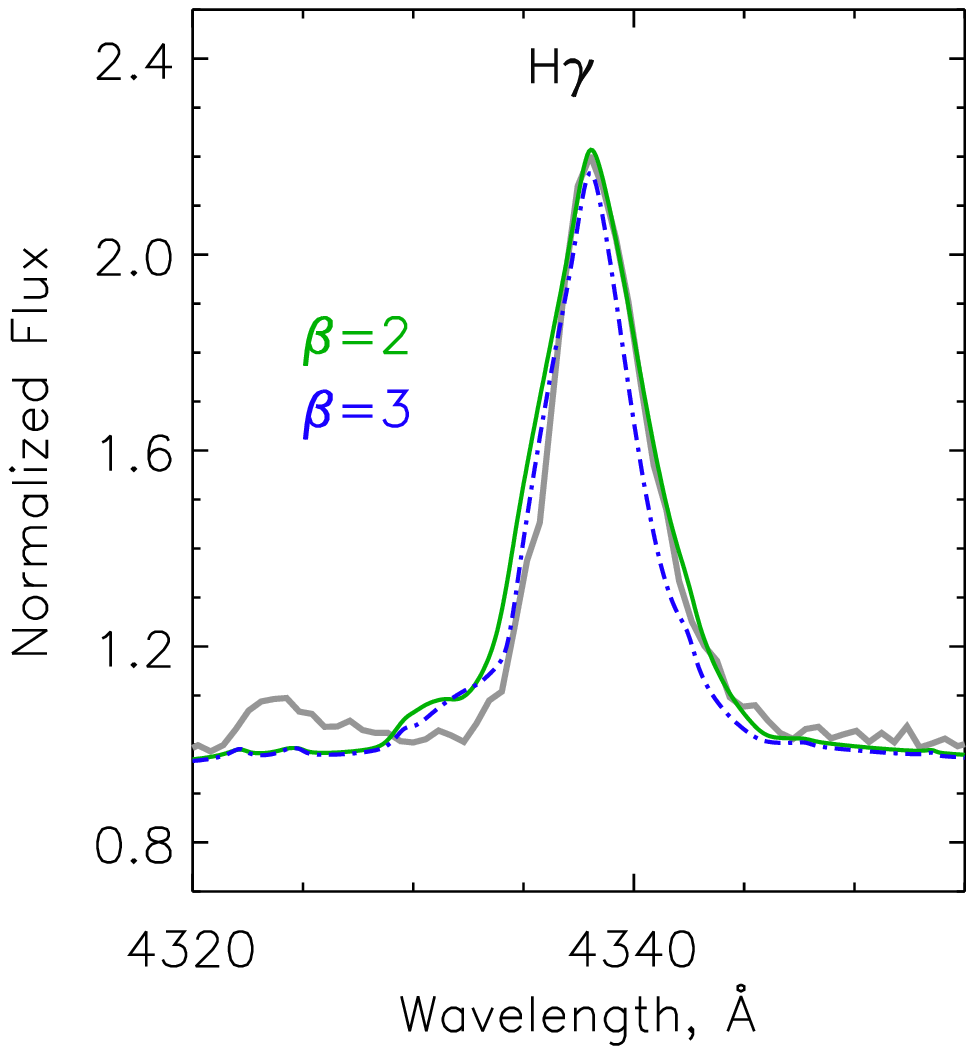}}}
{\centering \resizebox*{0.325\columnwidth}{!}{\includegraphics[angle=0,viewport=35 12 335 314,clip]{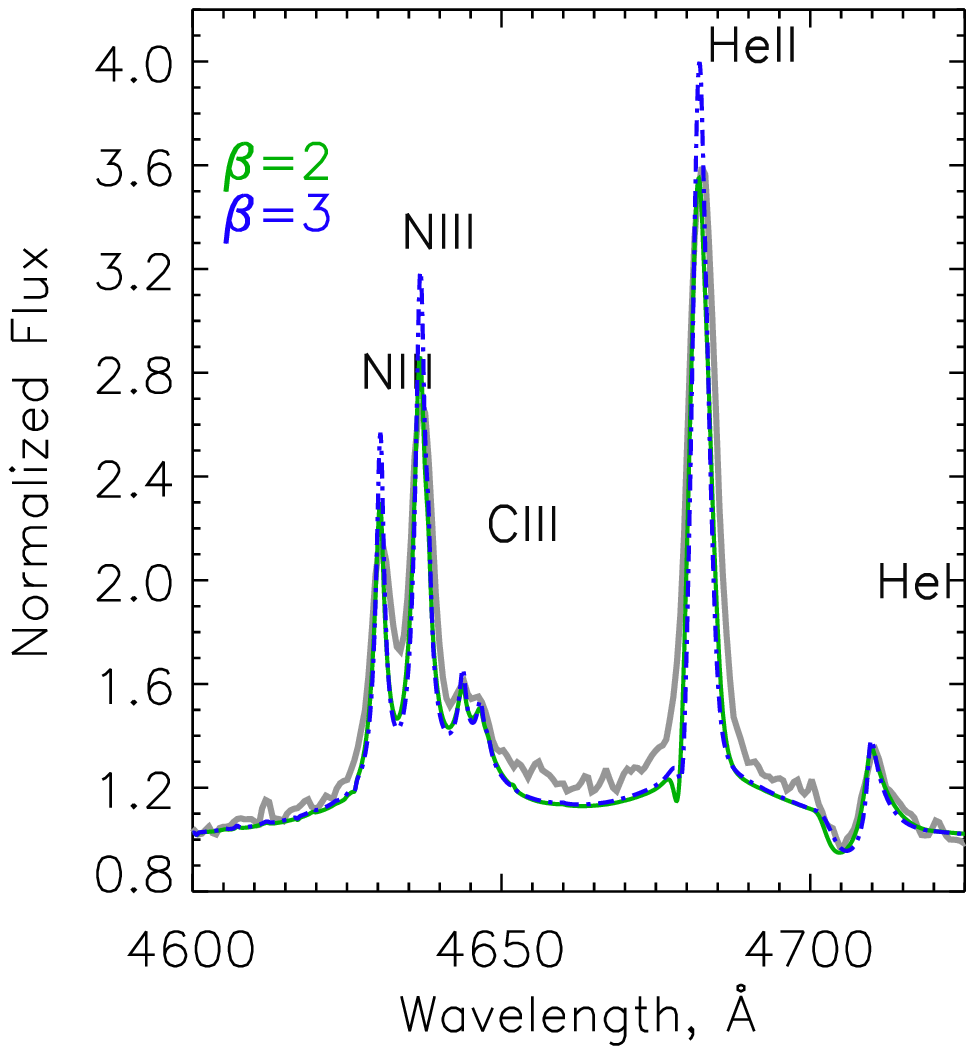}}}
{\centering \resizebox*{0.325\columnwidth}{!}{\includegraphics[angle=0,viewport=35 12 335 314,clip]{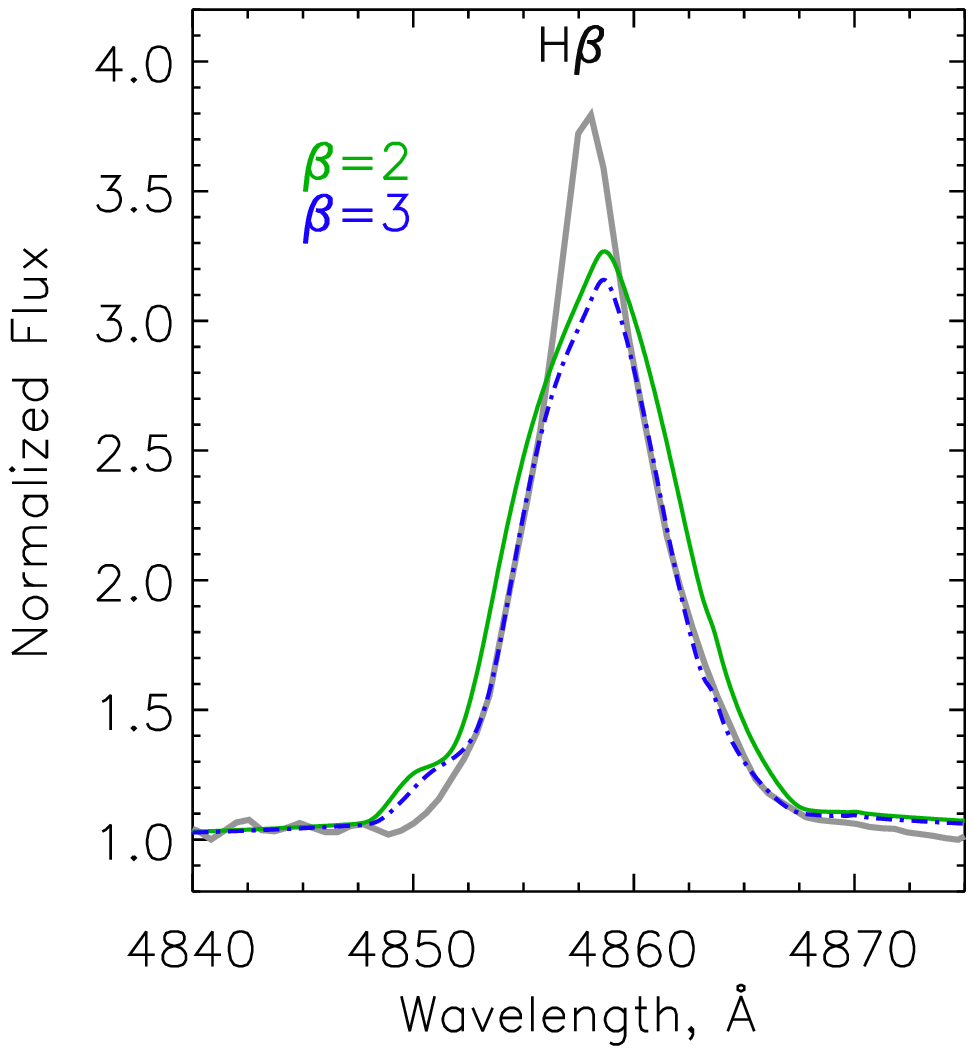}}}
{\centering \resizebox*{0.325\columnwidth}{!}{\includegraphics[angle=0,viewport=35 12 335 314,clip]{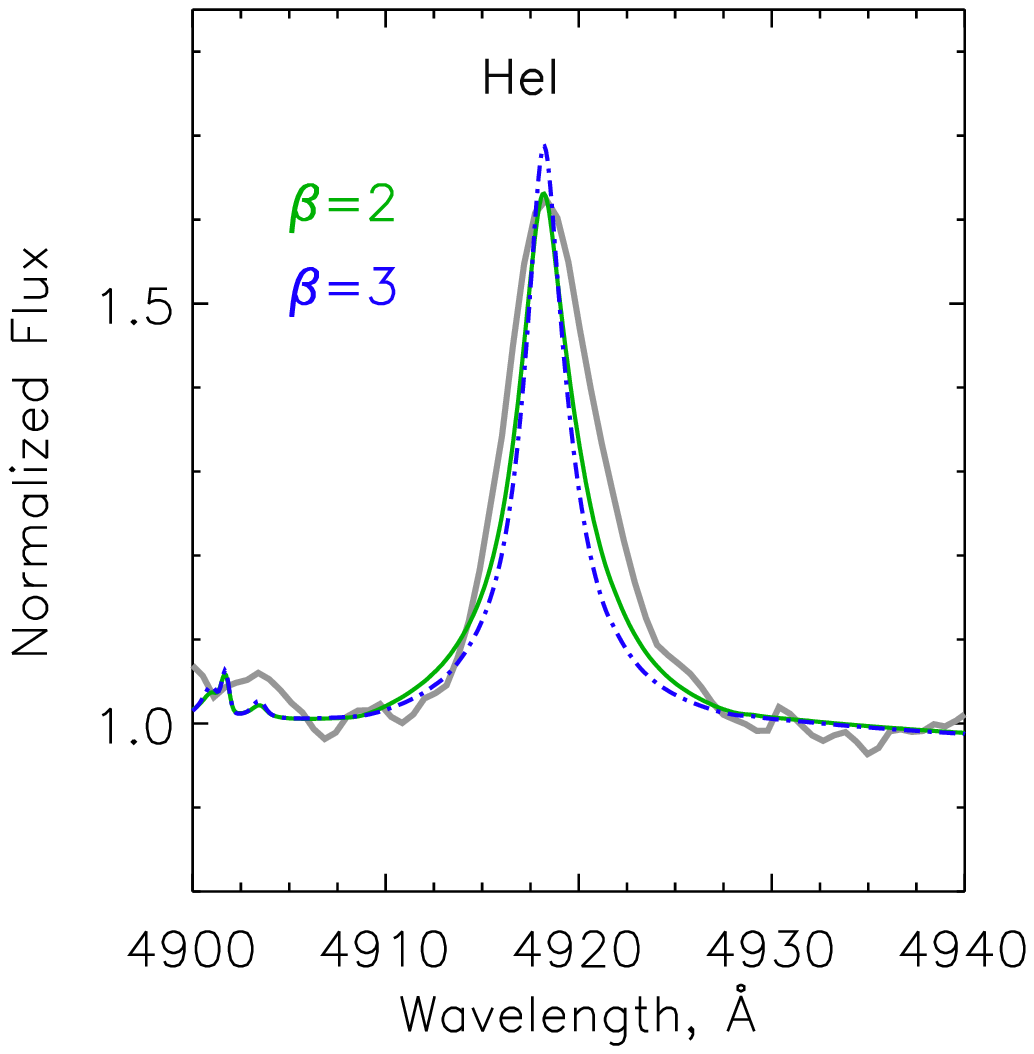}}}
{\centering \resizebox*{0.325\columnwidth}{!}{\includegraphics[angle=0,viewport=35 12 335 314,clip]{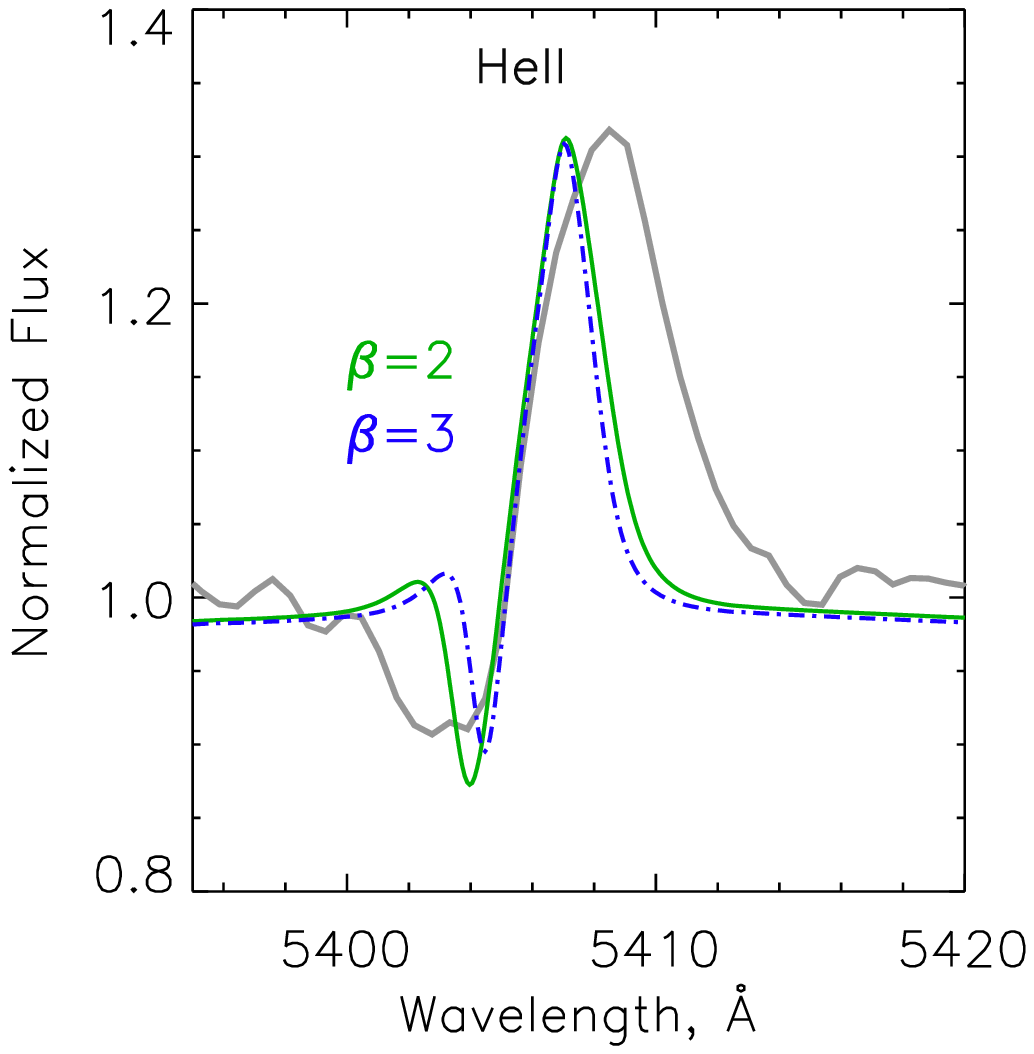}}}
{\centering \resizebox*{0.325\columnwidth}{!}{\includegraphics[angle=0,viewport=35 12 335 314,clip]{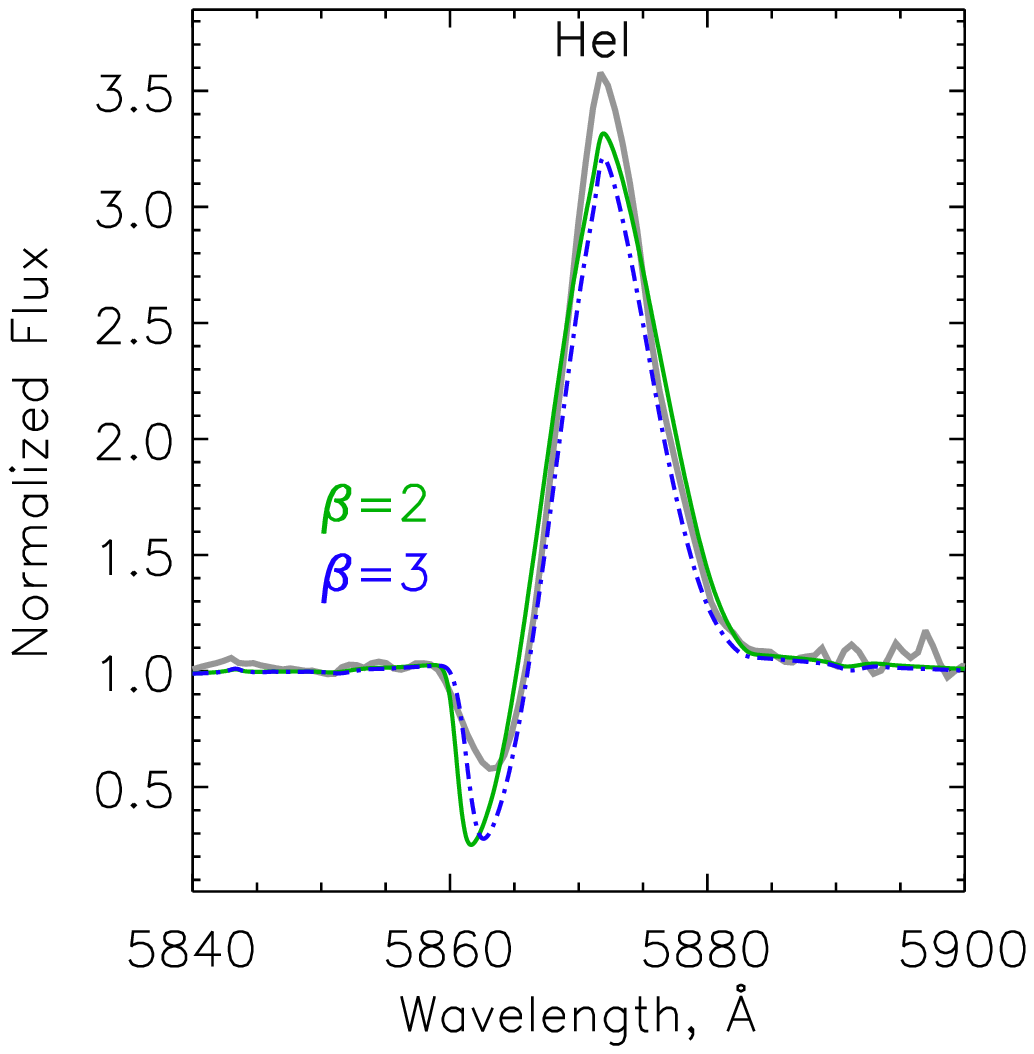}}}
{\centering \resizebox*{0.325\columnwidth}{!}{\includegraphics[angle=0,viewport=35 12 335 314,clip]{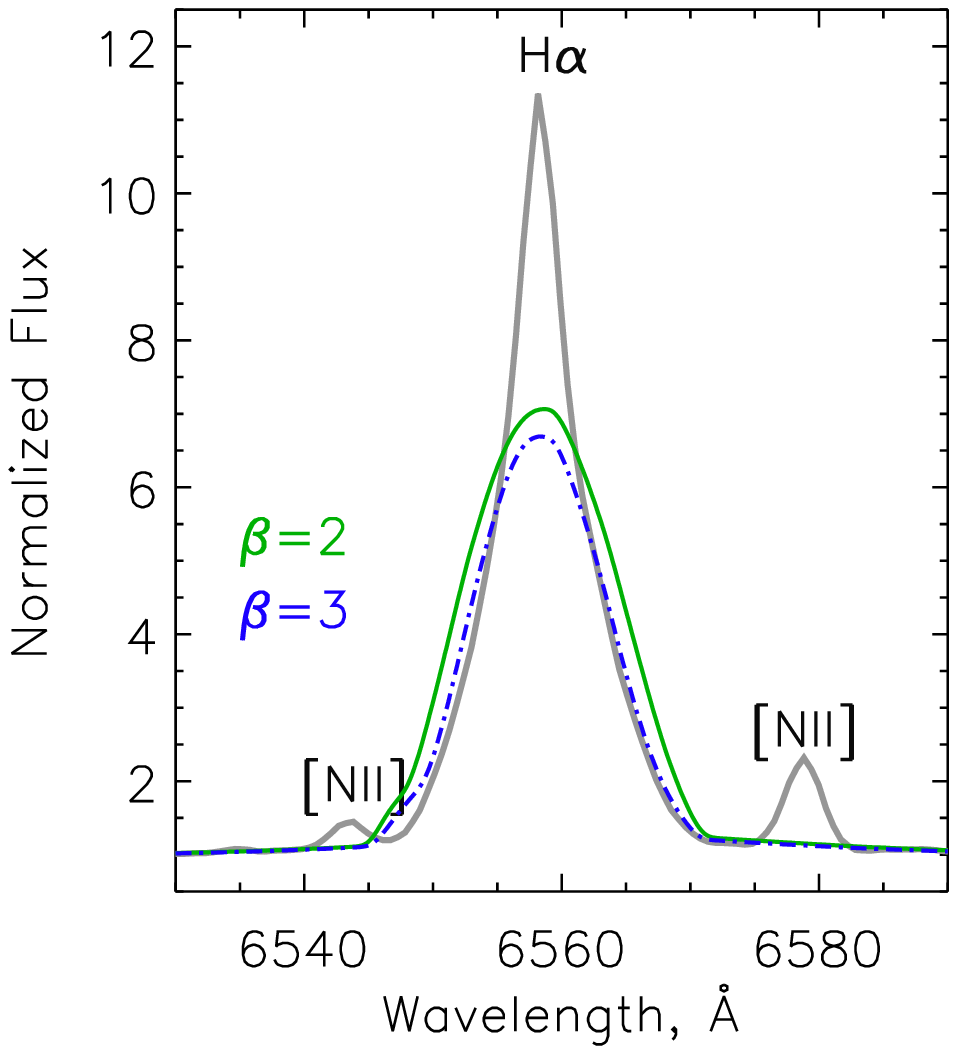}}}
{\centering \resizebox*{0.325\columnwidth}{!}{\includegraphics[angle=0,viewport=35 12 335 314,clip]{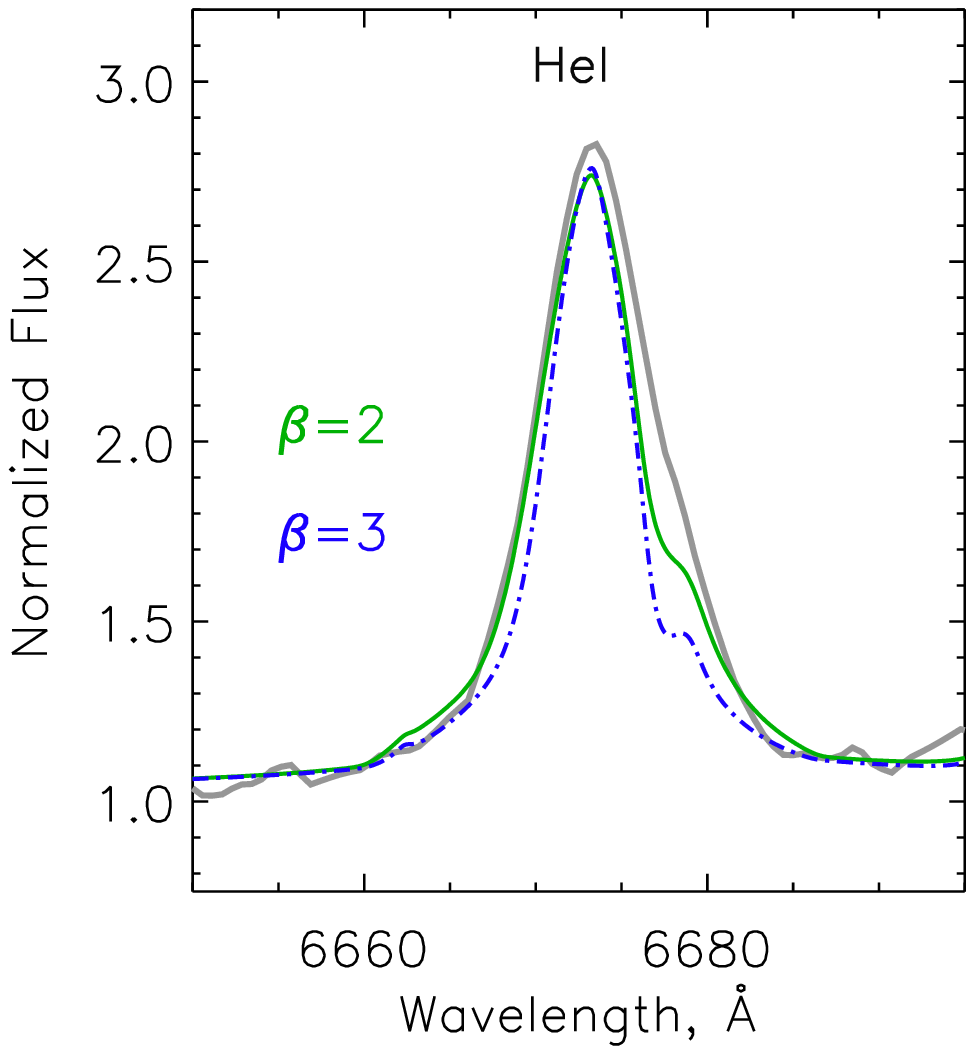}}}
\caption{Comparison of the profiles of selected lines with the best model spectra.}
\label{fig:modelsspectra2}
\end{figure*}

   The models with $\beta=2$ and $\beta=3$ are closest to reproducing the observed spectrum. Both models provide 
   a satisfactory fit to the overall spectral energy distribution and results of multicolor photometry as seen in Fig.~\ref{fig:model_sed}, 
   where the apparent discrepancy to the right of $\sim$6300\,\AA\ is caused by the uncertainties in the flux calibration 
   and the adopted reddening correction, as well as the strong emission line here. The detailed comparison of observed spectrum 
   with the best model with $\beta=2$ is shown in Fig.~\ref{fig:modelsspectra}. The Fig.~\ref{fig:modelsspectra2} clearly 
   demonstrates that some lines are better described with $\beta=2$, while others -- with $\beta=3$.

   The models coincide relatively well with the broad portion of the H$\alpha$, H$\beta$ profiles, lying 
   below the observed line maximum. This is consistent with the fact that GR\,290 lies within a nebular region
   from which significant H-Balmer emission arises and which is responsible for the sharp superposed emission
   in these lines (see Sect.~\ref{nebula} below). The emission component of the \ion{He}{i} P\,Cyg lines is 
   well-reproduced by the models, but not so the absorption component which tends to be weaker than predicted.  
   None of the models reproduces the profile shape of the high-excitation \ion{He}{II} 5411 \AA\ line,
   and this constitutes the largest discrepancy between the models and the observation. The model emission-line 
   profile is a factor of $\sim$2 narrower and the absorption component is significantly less extended than the 
   observations. Because of its high excitation, this line generally originates close to the hydrostatic 
   photosphere where the wind is still accelerating. This discrepancy suggests that the emission-line 
   forming region in GR\,290 may be more complex than assumed in the model. That is, it may not be spherically
   symmetrical, or portions of the wind emission may be dominated by shock emission, factors which are not
   incorporated in the current CMFGEN model. 

\section{The nebula surrounding GR\,290 \label{nebula}}

   Circumstellar gaseous nebulae are formed around massive stars during BSG \citep{Lamers2001} and LBV stages \citep{Weis2012ASPC}. 
   The sizes of these nebulae significantly increase under effect of fast stellar wind when the central stars transit to WR stage. 
   Nebulae can tell us the history of its stars' formation, ages, masses of ejected material. The presence of 
   forbidden lines in the spectra of GR\,290 indicates that it has a nebula, but the large distance prevents its detection on direct images. 
   Below we investigate the structure of GR\,290's nebula using its two-dimensional spectrum.

   The two-dimensional spectral image obtained with GTC contains information on the interstellar
   medium in the neighborhood of GR\,290. The long OSIRIS slit was oriented in the East-West direction
   and in addition to GR\,290, crosses the nearby OB88 and OB89 associations \citep{HumphreysSandage, Ivanov, MasseyArmandroff1995}.  
   OB89 and OB88 lie west of GR\,290. 
   
\begin{figure}
{\centering \resizebox*{0.95\columnwidth}{!}{\includegraphics[angle=0,viewport=0 0 302 214,clip]{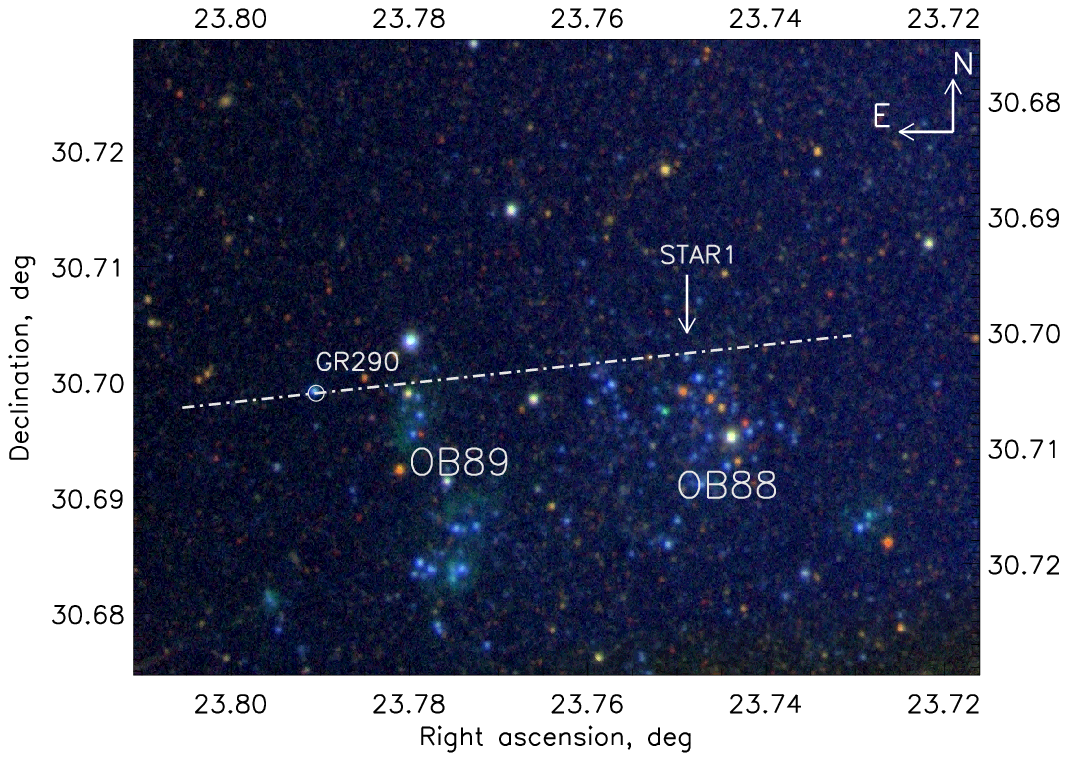}}}\\
{\centering \resizebox*{0.95\columnwidth}{!}{\includegraphics[angle=0,viewport=0 0 340 345,clip]{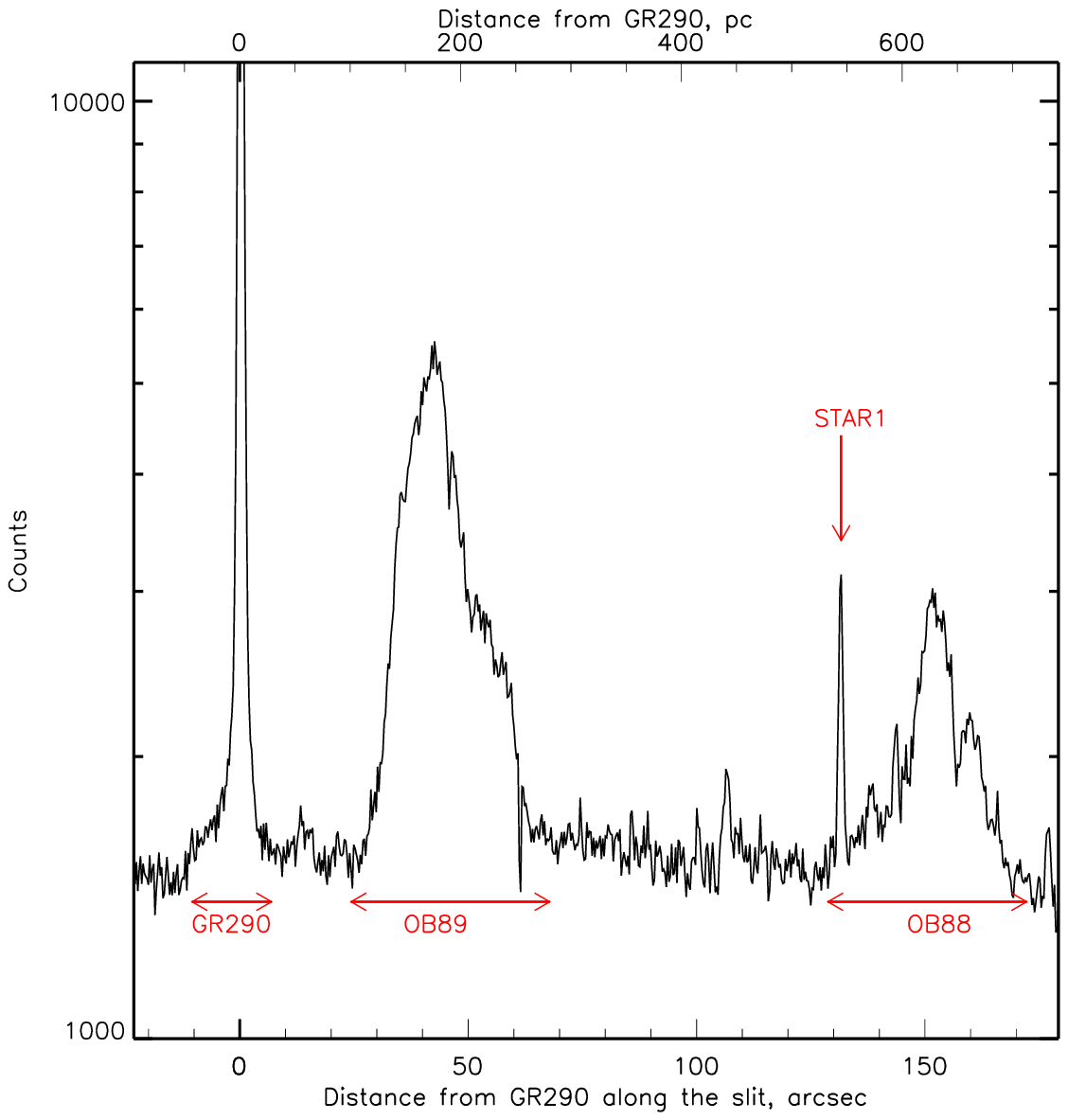}}}
\caption{Upper panel: identification chart for GR290. Dash-dotted line shows the slit position during July observations. 
Lower panel: tracing of the 2-dimensional spectral image obtained of GR\,290 in July 2017 along the
         spatial direction centred on the H$\alpha$ emission line. GR\,290 is located at zero point 
         and has a strong and sharply-peaked H$\alpha$ line with asymmetrically extended wings.          
\label{tracing_nebul_prof}} 
\end{figure}

   Figure~\ref{tracing_nebul_prof} illustrates the tracing in the spatial dimension along the H$\alpha$ emission line.  
   GR290 is at the origin of coordinates and appears sharp and narrow, compared to the other features in this tracing. 
   From an analogous tracing at [\ion{N}{ii}] 6548 \AA\, the spatial resolution is estimated to be $\sim$4 pixels, which is   
   approximately 4 pc assuming a M\,33 distance of 847 kpc. This is consistent with the limit imposed   
   by the ``seeing'' of the observations: $\sim$1.2'' in July and $\sim$1.6'' in August.  
   
   A Gaussian fit to the sharp stellar emission peak indicates that there is considerable H$\alpha$ emission 
   excess outside the wings of this fit, which can be associated with interstellar (ISM) emission. 
   Specifically, the H$\alpha$ emission appears to extend for 9 pixels\footnote{The ``plate scale'' 
   is approximately 0.29 arcsec/pix or, assuming $D_{M\,33}$=847 kpc,  1.2 pc/pixel} on the eastern side of the star 
   but only for $\sim$4 pixels on the western side. 

\begin{table}
\begin{center}
\begin{small}
\caption{Heliocentric radial velocities of nebular lines ($\rm{km~s^{-1}}$) August spectrum. \label{table_ISM}}
\begin{tabular}{lcccccc}
\hline
\hline
{\it Position:}                   &   1     &    2       & 3     \\
Line                              &{GR\,290}& {M\,33-back} &{OB89} \\
\hline
\hline
 $[\ion{S}{II}] \ 6716.44$        &...      &-208        &-223     \\  
 $[\ion{S}{II}] \ 6730.82$        &...      & ...        &-226     \\  
 $[\ion{N}{II}] \ 6548.01$        &-188     &-147:       &-180:$^a$\\
 $[\ion{N}{II}] \ 6583.45$        &-189     &-233        &-211     \\
 $H\alpha \ 6562.83$              &-179     &-194        &-200     \\
 $[\ion{O}{III}] \  4958.90$      &-206     & ...        & ...     \\
 $[\ion{O}{III}] \  5006.84$      &-201$^b$ & ...        & ...     \\
\hline
\hline
\end{tabular}
\end{small}
\end{center}
{\bf Notes:} $^a$ The [\ion{N}{ii}]6548 \AA\ line is significantly weaker than [\ion{N}{ii}]6583 and is partially blended with the $H\alpha$ emission line.  
             For this reason, we consider this measurement to have a larger uncertainty in RV than the rest, which is indicated with a colon (:) in the table; 
             $^b$ [\ion{O}{iii}]5007 was measured on the one-dimensional spectrum with a three-Gaussian fit
             to correct for the presence of the neighboring \ion{He}{i}5015 \AA\ line whose P\,Cyg absorption component
             overlaps its red wing. 
\end{table}

    The RVs of the more prominent ISM lines in the two-dimensional image are listed in Table~\ref{table_ISM}. 
    The measurements were performed at three locations along the spatial resolution axis. {\it Position 1:}  
    centered on GR\,290; {\it Position 2:} on the background close to GR\,290, which provides information 
    on the diffuse component of M\,33 in the neighborhood of GR\,290; and {\it Position 3:} centered on the 
    OB89 cluster. Column 1 lists the principal emission line identification with its corresponding 
    laboratory wavelength (\AA); and columns 2-4 list the heliocentric RVs of the lines at {\it Positions 1}, 
    {2} and {3}, respectively. The averages ($\pm$ s.d.) of the lines at the three positions are, 
    respectively -195 $\pm$14, -196$\pm$36 and -208$\pm$19 ${\rm km~s^{-1}}$, all consistent within the 
    uncertainties with each other and with the systemic speed of M\,33. The [\ion{N}{ii}]6548 is quite extended 
    and weak in the M\,33-background surrounding GR\,290 rendering its RV more uncertain than the rest. If 
    we eliminate this line, the average RV for this region is -211$\pm$20 ${\rm km~s^{-1}}$.

    H$\alpha$ is present at all three positions. The [\ion{S}{ii}] 6717, 6731 \AA\  lines are absent at {\it Position 1} 
    and the [\ion{O}{iii}] 4959, 5007 \AA\ lines are absent at {\it Positions 2} and {3}. \ion{He}{i} is 
    not detected at {\it Position 2}. The spectrum at {\it Position 1} also displays lines of [\ion{Fe}{iii}]5270 and [\ion{Ar}{iii}]7136. 
  
\vskip0.5cm

     We used the CLOUDY (version 17.00) photoionization code \cite{Ferland1998, Ferland2017} for revealing the 
     properties of the nebula surrounding the GR\,290 star. More than a thousand models were computed varying 
     chemical composition, density and outer radius of the nebula. The parameters of our best CMFGEN model were 
     used to set the luminosity, $T_{\rm eff}$ and spectral energy distribution of the central ionizing source. We 
     adopted a closed spherical geometry of the nebula with the same filling factor as for the star atmosphere.
     
     Two scenarios for the nebular chemical composition were considered. It could be formed from the surrounding 
     material, or from the LBV ejecta. For the first scenario we assumed that the chemical composition of the 
     compact ISM region is as usually found for \ion{H}{II} regions, scaled to the metallicity of M\,33 at the 
     galactocentric distance of GR\,290 ($\mathrm{12+\log(O/H) \simeq 8.15}$, or $Z \simeq 0.3Z_\odot$, 
     given the metallicity gradient derived by \cite{Rosolowsky2008, Bresolin2011}). 
     We also computed the models for 0.1, 0.2, 0.4 and 0.5 $Z_\odot$. In the second scenario we assumed that
     the compact nebula was formed primarily from LBV ejecta, and we adopted the chemical composition deduced 
     from the CMFGEN model (Table~\ref{table_abundances}). 
     
     For each set of abundances, the hydrogen density was varied in the range $\log(n_H) = 1.0 - 3.6$ and the 
     outer radius was varied in the $R=0.5 - 6$ pc. We also considered larger densities and radii, but the 
     corresponding models failed completely to fit the observed spectrum.
     
     The aim of the modeling was to reproduce the observed nebular emission lines that are clearly seen in 
     observed spectrum and not predicted by the CMFGEN model ([\ion{N}{ii}]6548, 6584; [\ion{O}{iii}]4959, 5007; 
     [\ion{Fe}{iii}]5270). These lines are assumed to arise in the the circumstellar medium that is  
     photoionized by GR\,290. An additional constraint on the model is the absence of the [\ion{S}{ii}]6717, 6731 \AA.
     The observed and modeled fluxes of these lines were compared and a chi-square value for each model was
     computed. The sum of CMFGEN model and  CLOUDY model spectra is presented in  Fig.~\ref{fig:modelsspectra}. 
     
\begin{figure}
        \includegraphics[width=\linewidth]{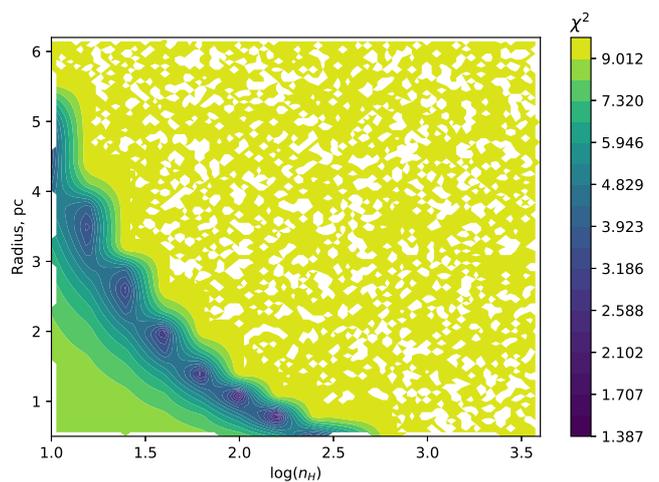}
        \caption{Chi-square distribution for the CLOUDY models with different densities and outer radii. 
        Chemical composition here is the same as for GR\,290 best-fit CMFGEN model, but with two times 
        larger abundance of O and Fe. For clarity all values of $\chi^2 > 10$ were changed to 10 in this plot.}\label{fig:CLOUDY_chi2}
\end{figure}

     We find that models for a nebula having the same chemical composition as the LBV stellar atmosphere yield  
     chi-square values that are several times smaller than the models in which the alternative set of abundances
     are used. This result favors the hypothesis that the nebula around GR\,290 has been formed primarily from material 
     ejected from the star. Another finding is that only relatively small outer radii yield spectra similar to 
     that observed: for $R > 4$ pc the nebular emission lines become much brighter than observed, at any density 
     and chemical abundance. This effect is clearly seen in Fig.~\ref{fig:CLOUDY_chi2}, where the chi-square 
     distribution for different densities and radii is shown and where the abundances are held constant.
     Although the values of the chi-squared distribution may vary for different sets of abundances, the general trend 
     is the same. 
     
     The minimum chi-square from all the computed models is obtained for an outer radius 
     $R=0.8$ pc and a hydrogen density $n_\mathrm{H} = 160\ \mathrm{cm^{-3}}$.  
      Unfortunately, the spatial resolution of our observations is limited by the ``seeing'' ($\sim$1''),
     which at the distance of M\,33 implies that we cannot resolve structures smaller than 4 pc in extension.

  All the computed models reproduce approximately the observations if the radius of this nebula does 
  not exceed 4 pc, with a preferred diameter being 2 pc. A larger nebula would produce brighter emission 
  lines than observed, regardless of density and metallicity.
  
  While most of the lines in the models are consistent with the observations, the [\ion{O}{iii}]4959, 5007 
  ones are 2-3 times fainter than observed if we assume that N/O=15 like in the atmosphere of the star. Moreover, such 
  high N/O ratio is inconsistent with the ones for other LBVs (Table~\ref{table_LBVnebulae}). This inconsistency could be resolved if the
  oxygen abundance of the nebula were 3 times greater than the one implied by the CMFGEN model -- i.e. if it is N/O=5 as we list 
  in Table~\ref{table_LBVnebulae}. The difference between chemical compositions of the nebula and the atmosphere does not 
  contradict the predictions of evolutionary theory. Indeed, according to the predictions of evolutionary tracks 
  (e.g. \citet{Groh2014evolution}) the surface abundance of oxygen rapidly drops by about 1.5 orders of magnitude during the LBV phase, while the nitrogen 
  abundance increases by only two times. At the same time, the nebula may be formed well before LBV stage, during the transition from main sequence 
  towards blue supergiants \citep{Lamers2001}, and therefore may have increased nitrogen and normal (equal to surrounding metallicity) oxygen abundance.

  The CLOUDY model only partially recovered the broad [\ion{Fe}{iii}]5270 presented in the spectrum of GR\,290 (Fig.~\ref{fig:modelsspectra}). 
  Probably it is related to the adopted geometry of the model that would require future investigations.

\begin{table*}
\begin{center}
\begin{small}
\caption{Nebular parameters for some LBVs  and of GR290. \label{table_LBVnebulae}}
\begin{tabular}{rl lll lll l}
\hline
\hline
\\
{\bf  Name }        &{\bf  Galaxy}& {\bf Type }  &{\bf Max.size}      &{\bf ${n_e}$ }                      & {\bf N/O}          &{\bf N/S}&  {\bf (N+O)/H }        & {\bf Ref}\\   
                    &             &              &{\bf (pc)}          &{\bf ($\mathrm{cm^{-3}}$)}          &                    &         &                        &          \\
\hline                                                                                                                              
\hline                                                                                                                              
\\                                                                                                                                  
AG\,Car              &   MW  & bipolar            & $1.4\times2$       &  $820\pm170$                       &  $ 6\pm2$          &   <45   &    $4\times10^{-4} $  & 1,\,2  \\
P\,Cyg               &   MW  & spherical          & 0.2/0.84           &  600                               &   >0.16            &    33   & $ 0.4-3\times10^{-3}$ & 1,\,2  \\
WRA\,751             &   MW  & spherical/bipolar  & 0.6                &  200                               & $3^{+3}_{-1.5}$    &         & $ 4-4.9\times10^{-4}$ & 1,\,3  \\
R\,127               &   LMC & bipolar            & 1.3                &  $720\pm90$                        &   $0.9\pm0.4$      &   <34   & $ 3.2\times10^{-4}$   & 1,\,2  \\
S\,119               &   LMC & spherical/outflow  & 1.8                &  $680\pm170$                       &   $1.9\pm0.5$      &   <78   & $ 3.8\times10^{-4}$   & 1,\,2  \\
\\                                                                                                                                     
\\                                                                                                                                     
GR\,290             &   M\,33 &                  &  0.8-4             &   $160^a$                          &   $~5^b$           & 65      & $7.3\times10^{-4} $   & this work \\  
\hline
\hline
\end{tabular}
\end{small}
\end{center}
{\bf Notes:} Types and sizes are taken from (1) \citet{Weis2011} while $n_e$ and chemical composition are taken from (2) \citet{Lamers2001} and (3) \citet{GarciaLario1998}. $^a$ Here we show CLOUDY-estimated $n_H$ as $n_e$. $^b$ See the text for the discussion on how we estimated this quantity.
\end{table*}

    Table~\ref{table_LBVnebulae} presents the comparison of GR\,290's nebula parameters with the ones of four other LBVs belonging to our Galaxy~(MW) and Large Magellanic Cloud~(LMC). 
    As we can see the sizes of all five nebulae are similar, and the one of GR\,290's nebula is in good agreement with average sizes of LBVs according to \citet{Weis2012ASPC}.  
    However, $n_e$ for GR\,290's nebula is significantly lower\footnote{Here we assume that $n_H$ is equivalent to $n_e$.}, just like the one for WRA\,751 \citep{GarciaLario1998}.


    \citet{fabrika} reported the marginal detection of a circumstellar nebula around GR\,290 based on 
    integral field spectroscopy.
    They found a radial velocity gradient within the nebula of $44 \pm 11$ ${\rm km~s^{-1}}$ and an angular size $\sim 9$''  
    in the NE-SW direction. The detection is classified as marginal because, their spectra did not cover the 
    H$\alpha$ spectral region and they did not detect the [\ion{O}{iii}] lines, so the results are based on 
    the analysis of the H$\beta$ line profiles alone. However, if real, this region would correspond to
    material lying intermediate between the compact \ion{H}{II} region that we detect (R$\leq$4 pc) in the 
    emission-line spectrum and the material that we are able to resolve at a projected distance of 
    16-40 pc to the East of GR\,290. If this intermediate material reflects an asymmetric ejection, it
    increases the possibility that the inner compact \ion{H}{II} region is also asymmetrical.
    It is also interesting to note that a preliminary CLOUDY photoionization model indicates that GR\,290 could
    be responsible for producing at least a fraction of the observed [\ion{S}{II}] at these distances, with 
    no [\ion{O}{iii}]. However, a more detailed analysis of the extended regions 
    requires observations with higher S/N than those we currently present.


\section{Conclusions}
     We present results of long-slit observations obtained with GTC in 2016 of the LBV/WR
     object GR\,290 (Romano's star). GR\,290 lies in a relatively external region of the 
     M\,33 galaxy, where metallicity is low compared to Solar values.
     Fitting CMFGEN models to the stellar spectrum we find a current effective
     temperature $T_{\rm eff}=27000-30000$ K at a radius $R_{2/3}=27-23 {\rm R_\odot}$.
     It's mass-loss rate is 1.5$\times$10$^{-5}$ ${\rm M_\odot yr^{-1}}$.  
     
     The terminal wind speed $\vv_{\infty}$=620 ${\rm km~s^{-1}}$ is faster than ever before
     recorded while the current luminosity $L_*=(3.1-3.7)\times$10$^5$ ${\rm L_\odot}$ is the lowest
     ever deduced. The remaining parameters of the 2016 observation fall within the ranges of
     values obtained at previous epochs as described in \citet{Polcaro2016}\footnote{ The larger $T_{\rm eff}$ derived by \citet{Polcaro2016}
     for GR290 during the 2008 and 2014 minima can be attributed to the fact that the stellar parameters have been computed adopting $\beta=1$.} 
     We also confirm previous conclusions that the stellar wind is overabundant in He and N and underabundant 
     in C and O.  
     
     We find the star in GR\,290 to be surrounded by an unresolved compact \ion{H}{II} region
     with dimensions $\leq$4 pc, and having chemical abundances that are consistent with those derived
     from the stellar wind lines. Hence, this compact \ion{H}{II} region appears to be largely composed of
     material ejected from the star. In addition, we find partially resolved emission from a more extended 
     ISM region which appears to be asymmetrical, with a larger extent to the East ($16-40$ pc) than to the West. 
     If associated with GR\,290, this more extended structure would suggest that GR\,290
     has undergone significant mass-loss over the past 10$^4$-10$^5$ years. 
     
     The luminosity and effective temperature of GR\,290 allow us to place it on the HRD
     and estimate its mass. Using the \citet{Brott2011} 
     non-rotating evolutionary grid, $M\sim 40-50~{\rm M_\odot}$, while with the \citet{ChieffiLimongi2013}
     tracks its mass is $\sim$60 ${\rm M_\odot}$. The enhanced He and N chemical abundances,
     along with deficiencies in C and O derived from the CMFGEN models indicate that the
     surface layers of GR\,290 contain significant amounts of nuclear processed material, such
     as would be expected for a star that has left the Main Sequence and is becoming a WR star
     \citep{Langer1994}. Adding this to the recent LBV-like eruptive behavior and to
     the presence of a compact \ion{H}{II} region also having altered chemical abundances with respecto
     to typical \ion{H}{II} regions, one is led to conclude that GR\,290 is in an advanced stage of evolution,
     either nearing the end of its core H-burning phase or in later phases.

     There are now a number of reported supernova (SNIIn) events that have been observed in
     extragalactic regions where there has been a massive ejection of material prior to the SN event
     -- often within 10 years -- and which is believed to have arisen in an LBV
     \citep{FoleySmith2007,SmithLi2007,SmithLiFilippenko2011,Smith2007SN1987,Pastorello2008,Dessart2011}.
     Thus, given the apparent late evolutionary state of GR\,290, it is a good candidate to be a
     progenitor of a similar event sometime in the future.

\section{Acknowledgements}

     We dedicate this publication to the memory of our dear colleague Vito Francesco Polcaro, who
     passed away shortly before the final acceptance of the paper.
     
     Dr. Vito Francesco Polcaro passed away on February 11, 2018 in Roma. Besides GR\,290, Dr. Polcaro prompted many relevant investigations of hot luminous stars,
     X-ray counterparts, and circumstellar nebulae, best exploiting in particular the performances of medium and small telescopes.
     He inspired amateurs to participate in the photometric monitoring of topical objects. Last years Dr. Polcaro has been very active in the field of archeoastronomy, 
     and organized campaigns in paleolithic sites and conferences gathering astronomers and archeologists. 
     
     \vskip0.5cm
     
     We would like to thank  the anonymous referee for a helpful and insightful report. 
     We thank the GTC observatory staff for obtaining the spectra and Antonio Cabrera-Lavers for guidance in processing 
     the observations. GK acknowledges support from CONACYT grant 252499 and thanks Ulises Amaya for computer support. 
     OM acknowledges Russian Foundation for Basic Research grant 16-02-00148 and AV \v{C}R project RVO:67985815. 
     OE acknowledges Russian Foundation for Basic Research grant 18-02-00976.

\bibliographystyle{aa}
\bibliography{GR290_GTC}

\begin{thebibliography}{66}
\expandafter\ifx\csname natexlab\endcsname\relax\def\natexlab#1{#1}\fi

\bibitem[{{Abolmasov}(2011)}]{PashaLBV}
{Abolmasov}, P. 2011, \na, 16, 421

\bibitem[{{Afanasiev} \& {Moiseev}(2005)}]{AfanasievMoiseev2005}
{Afanasiev}, V.~L. \& {Moiseev}, A.~V. 2005, Astronomy Letters, 31, 194

\bibitem[{{Bresolin}(2011)}]{Bresolin2011}
{Bresolin}, F. 2011, \apj, 730, 129

\bibitem[{{Brott} {et~al.}(2011){Brott}, {de Mink}, {Cantiello}, {Langer}, {de
  Koter}, {Evans}, {Hunter}, {Trundle}, \& {Vink}}]{Brott2011}
{Brott}, I., {de Mink}, S.~E., {Cantiello}, M., {et~al.} 2011, \aap, 530, A115

\bibitem[{{Chieffi} \& {Limongi}(2013)}]{ChieffiLimongi2013}
{Chieffi}, A. \& {Limongi}, M. 2013, \apj, 764, 21

\bibitem[{{Clark} {et~al.}(2012){Clark}, {Castro}, {Garcia}, {Herrero},
  {Najarro}, {Negueruela}, {Ritchie}, \& {Smith}}]{ClarkLBV2012}
{Clark}, J.~S., {Castro}, N., {Garcia}, M., {et~al.} 2012, \aap, 541, A146

\bibitem[{{Conti}(1984)}]{Conti}
{Conti}, P.~S. 1984, in IAU Symposium, Vol. 105, Observational Tests of the
  Stellar Evolution Theory, ed. A.~{Maeder} \& A.~{Renzini}, 233

\bibitem[{{Dessart} {et~al.}(2011){Dessart}, {Hillier}, {Livne}, {Yoon},
  {Woosley}, {Waldman}, \& {Langer}}]{Dessart2011}
{Dessart}, L., {Hillier}, D.~J., {Livne}, E., {et~al.} 2011, \mnras, 414, 2985

\bibitem[{{Ekstr{\"o}m} {et~al.}(2012){Ekstr{\"o}m}, {Georgy}, {Eggenberger},
  {Meynet}, {Mowlavi}, {Wyttenbach}, {Granada}, {Decressin}, {Hirschi},
  {Frischknecht}, {Charbonnel}, \& {Maeder}}]{Ekstrom}
{Ekstr{\"o}m}, S., {Georgy}, C., {Eggenberger}, P., {et~al.} 2012, \aap, 537,
  A146

\bibitem[{{Fabrika} {et~al.}(2005){Fabrika}, {Sholukhova}, {Becker},
  {Afanasiev}, {Roth}, \& {Sanchez}}]{fabrika}
{Fabrika}, S., {Sholukhova}, O., {Becker}, T., {et~al.} 2005, \aap, 437, 217

\bibitem[{{Ferland} {et~al.}(2017){Ferland}, {Chatzikos}, {Guzm{\'a}n},
  {Lykins}, {van Hoof}, {Williams}, {Abel}, {Badnell}, {Keenan}, {Porter}, \&
  {Stancil}}]{Ferland2017}
{Ferland}, G.~J., {Chatzikos}, M., {Guzm{\'a}n}, F., {et~al.} 2017, Revista
  Mexicana de Astronomi\'a y Astrofi\'sica, 53, 385

\bibitem[{{Ferland} {et~al.}(1998){Ferland}, {Korista}, {Verner}, {Ferguson},
  {Kingdon}, \& {Verner}}]{Ferland1998}
{Ferland}, G.~J., {Korista}, K.~T., {Verner}, D.~A., {et~al.} 1998, \pasp, 110,
  761

\bibitem[{{Foley} {et~al.}(2007){Foley}, {Smith}, {Ganeshalingam}, {Li},
  {Chornock}, \& {Filippenko}}]{FoleySmith2007}
{Foley}, R.~J., {Smith}, N., {Ganeshalingam}, M., {et~al.} 2007, \apjl, 657,
  L105

\bibitem[{{Galleti} {et~al.}(2004){Galleti}, {Bellazzini}, \&
  {Ferraro}}]{Galleti2004}
{Galleti}, S., {Bellazzini}, M., \& {Ferraro}, F.~R. 2004, \aap, 423, 925

\bibitem[{{Garcia-Lario} {et~al.}(1998){Garcia-Lario}, {Riera}, \&
  {Manchado}}]{GarciaLario1998}
{Garcia-Lario}, P., {Riera}, A., \& {Manchado}, A. 1998, \aap, 334, 1007

\bibitem[{{Georgiev} {et~al.}(2011){Georgiev}, {Koenigsberger}, {Hillier},
  {Morrell}, {Barb{\'a}}, \& {Gamen}}]{Georgiev2011}
{Georgiev}, L., {Koenigsberger}, G., {Hillier}, D.~J., {et~al.} 2011, \aj, 142,
  191

\bibitem[{{Gr{\"a}fener} {et~al.}(2012){Gr{\"a}fener}, {Owocki}, \&
  {Vink}}]{Grafener2012}
{Gr{\"a}fener}, G., {Owocki}, S.~P., \& {Vink}, J.~S. 2012, \aap, 538, A40

\bibitem[{{Groh} {et~al.}(2009){Groh}, {Hillier}, {Damineli}, {Whitelock},
  {Marang}, \& {Rossi}}]{Groh2009AGCar}
{Groh}, J.~H., {Hillier}, D.~J., {Damineli}, A., {et~al.} 2009, \apj, 698, 1698

\bibitem[{{Groh} {et~al.}(2013{\natexlab{a}}){Groh}, {Meynet}, \&
  {Ekstr{\"o}m}}]{GrohMeynet2013}
{Groh}, J.~H., {Meynet}, G., \& {Ekstr{\"o}m}, S. 2013{\natexlab{a}}, \aap,
  550, L7

\bibitem[{{Groh} {et~al.}(2014){Groh}, {Meynet}, {Ekstr{\"o}m}, \&
  {Georgy}}]{Groh2014evolution}
{Groh}, J.~H., {Meynet}, G., {Ekstr{\"o}m}, S., \& {Georgy}, C. 2014, \aap,
  564, A30

\bibitem[{{Groh} {et~al.}(2013{\natexlab{b}}){Groh}, {Meynet}, {Georgy}, \&
  {Ekstr{\"o}m}}]{Groh2013GRB}
{Groh}, J.~H., {Meynet}, G., {Georgy}, C., \& {Ekstr{\"o}m}, S.
  2013{\natexlab{b}}, \aap, 558, A131

\bibitem[{{Hillier} \& {Miller}(1998)}]{Hillier5}
{Hillier}, D.~J. \& {Miller}, D.~L. 1998, \apj, 496, 407

\bibitem[{{Hillier} \& {Miller}(1999)}]{HillierMiller1999}
{Hillier}, D.~J. \& {Miller}, D.~L. 1999, \apj, 519, 354

\bibitem[{{Humphreys} \& {Davidson}(1994)}]{HumphreysDavidson}
{Humphreys}, R.~M. \& {Davidson}, K. 1994, \pasp, 106, 1025

\bibitem[{{Humphreys} \& {Sandage}(1980)}]{HumphreysSandage}
{Humphreys}, R.~M. \& {Sandage}, A. 1980, \apjs, 44, 319

\bibitem[{{Humphreys} {et~al.}(2014){Humphreys}, {Weis}, {Davidson}, {Bomans},
  \& {Burggraf}}]{Humphreys2014}
{Humphreys}, R.~M., {Weis}, K., {Davidson}, K., {Bomans}, D.~J., \& {Burggraf},
  B. 2014, \apj, 790, 48

\bibitem[{{Humphreys} {et~al.}(2016){Humphreys}, {Weis}, {Davidson}, \&
  {Gordon}}]{Humphreys2016}
{Humphreys}, R.~M., {Weis}, K., {Davidson}, K., \& {Gordon}, M.~S. 2016, \apj,
  825, 64

\bibitem[{{Ivanov} {et~al.}(1993){Ivanov}, {Freedman}, \& {Madore}}]{Ivanov}
{Ivanov}, G.~R., {Freedman}, W.~L., \& {Madore}, B.~F. 1993, \apjs, 89, 85

\bibitem[{{Koenigsberger} {et~al.}(2014){Koenigsberger}, {Morrell}, {Hillier},
  {Gamen}, {Schneider}, {Gonz{\'a}lez-Jim{\'e}nez}, {Langer}, \&
  {Barb{\'a}}}]{KoenigsbergerMorrell2014}
{Koenigsberger}, G., {Morrell}, N., {Hillier}, D.~J., {et~al.} 2014, \aj, 148,
  62

\bibitem[{{Kurtev} {et~al.}(2001){Kurtev}, {Sholukhova}, {Borissova}, \&
  {Georgiev}}]{Kurtev2001}
{Kurtev}, R., {Sholukhova}, O., {Borissova}, J., \& {Georgiev}, L. 2001,
  \rmxaa, 37, 57

\bibitem[{{Lamers} \& {Cassinelli}(1999)}]{LamersCassinelliBook}
{Lamers}, H.~J.~G.~L.~M. \& {Cassinelli}, J.~P. 1999, {Introduction to Stellar
  Winds}, 452

\bibitem[{{Lamers} {et~al.}(2001){Lamers}, {Nota}, {Panagia}, {Smith}, \&
  {Langer}}]{Lamers2001}
{Lamers}, H.~J.~G.~L.~M., {Nota}, A., {Panagia}, N., {Smith}, L.~J., \&
  {Langer}, N. 2001, \apj, 551, 764

\bibitem[{{Langer} {et~al.}(1994){Langer}, {Hamann}, {Lennon}, {Najarro},
  {Pauldrach}, \& {Puls}}]{Langer1994}
{Langer}, N., {Hamann}, W.-R., {Lennon}, M., {et~al.} 1994, \aap, 290, 819

\bibitem[{{Magrini} {et~al.}(2010){Magrini}, {Stanghellini}, {Corbelli},
  {Galli}, \& {Villaver}}]{magri10}
{Magrini}, L., {Stanghellini}, L., {Corbelli}, E., {Galli}, D., \& {Villaver},
  E. 2010, \aap, 512, A63

\bibitem[{{Martayan} {et~al.}(2016){Martayan}, {Lobel}, {Baade}, {Mehner},
  {Rivinius}, {Boffin}, {Girard}, {Mawet}, {Montagnier}, {Blomme}, {Kervella},
  {Sana}, {{\v S}tefl}, {Zorec}, {Lacour}, {Le Bouquin}, {Martins},
  {M{\'e}rand}, {Patru}, {Selman}, \& {Fr{\'e}mat}}]{MartayanLobel2016}
{Martayan}, C., {Lobel}, A., {Baade}, D., {et~al.} 2016, \aap, 587, A115

\bibitem[{{Maryeva} \& {Abolmasov}(2010)}]{maryeva2010}
{Maryeva}, O. \& {Abolmasov}, P. 2010, \rmxaa, 46, 279

\bibitem[{{Maryeva} \& {Abolmasov}(2012)}]{maryeva2012}
{Maryeva}, O. \& {Abolmasov}, P. 2012, \mnras, 419, 1455

\bibitem[{{Massey} {et~al.}(1995){Massey}, {Armandroff}, {Pyke}, {Patel}, \&
  {Wilson}}]{MasseyArmandroff1995}
{Massey}, P., {Armandroff}, T.~E., {Pyke}, R., {Patel}, K., \& {Wilson}, C.~D.
  1995, \aj, 110, 2715

\bibitem[{{Meynet} {et~al.}(2011){Meynet}, {Georgy}, {Hirschi}, {Maeder},
  {Massey}, {Przybilla}, \& {Nieva}}]{Meynet2011}
{Meynet}, G., {Georgy}, C., {Hirschi}, R., {et~al.} 2011, Bulletin de la
  Societe Royale des Sciences de Liege, 80, 266

\bibitem[{{Meynet} \& {Maeder}(2005)}]{Meynet2005eolutionWR}
{Meynet}, G. \& {Maeder}, A. 2005, \aap, 429, 581

\bibitem[{{Pastorello} {et~al.}(2008){Pastorello}, {Quimby}, {Smartt},
  {Mattila}, {Navasardyan}, {Crockett}, {Elias-Rosa}, {Mondol}, {Wheeler}, \&
  {Young}}]{Pastorello2008}
{Pastorello}, A., {Quimby}, R.~M., {Smartt}, S.~J., {et~al.} 2008, \mnras, 389,
  131

\bibitem[{{Polcaro} {et~al.}(2003){Polcaro}, {Gualandi}, {Norci}, {Rossi}, \&
  {Viotti}}]{Polcaro2003}
{Polcaro}, V.~F., {Gualandi}, R., {Norci}, L., {Rossi}, C., \& {Viotti}, R.~F.
  2003, \aap, 411, 193

\bibitem[{{Polcaro} {et~al.}(2016){Polcaro}, {Maryeva}, {Nesci}, {Calabresi},
  {Chieffi}, {Galleti}, {Gualandi}, {Haver}, {Mills}, {Osborn}, {Pasquali},
  {Rossi}, {Vasilyeva}, \& {Viotti}}]{Polcaro2016}
{Polcaro}, V.~F., {Maryeva}, O., {Nesci}, R., {et~al.} 2016, \aj, 151, 149

\bibitem[{{Polcaro} {et~al.}(2011){Polcaro}, {Rossi}, {Viotti}, {Galleti},
  {Gualandi}, \& {Norci}}]{polcaro10}
{Polcaro}, V.~F., {Rossi}, C., {Viotti}, R.~F., {et~al.} 2011, \aj, 141, 18

\bibitem[{{Puls} {et~al.}(1996){Puls}, {Kudritzki}, {Herrero}, {Pauldrach},
  {Haser}, {Lennon}, {Gabler}, {Voels}, {Vilchez}, {Wachter}, \&
  {Feldmeier}}]{Puls1996}
{Puls}, J., {Kudritzki}, R.-P., {Herrero}, A., {et~al.} 1996, \aap, 305, 171

\bibitem[{{Romano}(1978)}]{romano}
{Romano}, G. 1978, \aap, 67, 291

\bibitem[{{Rosolowsky} \& {Simon}(2008)}]{Rosolowsky2008}
{Rosolowsky}, E. \& {Simon}, J.~D. 2008, \apj, 675, 1213

\bibitem[{{Sander} {et~al.}(2017){Sander}, {Hamann}, {Todt}, {Hainich}, \&
  {Shenar}}]{Sander2017beta}
{Sander}, A.~A.~C., {Hamann}, W.-R., {Todt}, H., {Hainich}, R., \& {Shenar}, T.
  2017, \aap, 603, A86

\bibitem[{{Sholukhova} {et~al.}(2002){Sholukhova}, {Zharova}, {Fabrika}, \&
  {Malinovskii}}]{Sholukhova2002}
{Sholukhova}, O., {Zharova}, A., {Fabrika}, S., \& {Malinovskii}, D. 2002, in
  Astronomical Society of the Pacific Conference Series, Vol. 259, IAU Colloq.
  185: Radial and Nonradial Pulsationsn as Probes of Stellar Physics, ed.
  C.~{Aerts}, T.~R. {Bedding}, \& J.~{Christensen-Dalsgaard}, 522

\bibitem[{{Sholukhova} {et~al.}(1997){Sholukhova}, {Fabrika}, {Vlasyuk}, \&
  {Burenkov}}]{Sholukhova97}
{Sholukhova}, O.~N., {Fabrika}, S.~N., {Vlasyuk}, V.~V., \& {Burenkov}, A.~N.
  1997, Astronomy Letters, 23, 458

\bibitem[{{Sholukhova} {et~al.}(2011){Sholukhova}, {Fabrika}, {Zharova},
  {Valeev}, \& {Goranskij}}]{Sholukhova2011}
{Sholukhova}, O.~N., {Fabrika}, S.~N., {Zharova}, A.~V., {Valeev}, A.~F., \&
  {Goranskij}, V.~P. 2011, Astrophysical Bulletin, 66, 123

\bibitem[{{Smith}(1968)}]{Smith1968}
{Smith}, L.~F. 1968, \mnras, 138, 109

\bibitem[{{Smith} {et~al.}(1996){Smith}, {Shara}, \&
  {Moffat}}]{SmithSharaMoffat1996}
{Smith}, L.~F., {Shara}, M.~M., \& {Moffat}, A.~F.~J. 1996, \mnras, 281, 163

\bibitem[{{Smith}(2007)}]{Smith2007SN1987}
{Smith}, N. 2007, in American Institute of Physics Conference Series, Vol. 937,
  Supernova 1987A: 20 Years After: Supernovae and Gamma-Ray Bursters, ed.
  S.~{Immler}, K.~{Weiler}, \& R.~{McCray}, 163--170

\bibitem[{{Smith} {et~al.}(2011){Smith}, {Li}, {Filippenko}, \&
  {Chornock}}]{SmithLiFilippenko2011}
{Smith}, N., {Li}, W., {Filippenko}, A.~V., \& {Chornock}, R. 2011, \mnras,
  412, 1522

\bibitem[{{Smith} {et~al.}(2007){Smith}, {Li}, {Foley}, {Wheeler}, {Pooley},
  {Chornock}, {Filippenko}, {Silverman}, {Quimby}, {Bloom}, \&
  {Hansen}}]{SmithLi2007}
{Smith}, N., {Li}, W., {Foley}, R.~J., {et~al.} 2007, \apj, 666, 1116

\bibitem[{{Szeifert}(1996)}]{szeifert}
{Szeifert}, T. 1996, in Liege International Astrophysical Colloquia, Vol.~33,
  Liege International Astrophysical Colloquia, ed. J.~M. {Vreux}, A.~{Detal},
  D.~{Fraipont-Caro}, E.~{Gosset}, \& G.~{Rauw}, 459

\bibitem[{{van der Hucht} {et~al.}(1981){van der Hucht}, {Conti}, {Lundstrom},
  \& {Stenholm}}]{vanderHucht1981}
{van der Hucht}, K.~A., {Conti}, P.~S., {Lundstrom}, I., \& {Stenholm}, B.
  1981, \ssr, 28, 227

\bibitem[{{Vink}(2012)}]{Vink2012review}
{Vink}, J.~S. 2012, in Astrophysics and Space Science Library, Vol. 384, Eta
  Carinae and the Supernova Impostors, ed. K.~{Davidson} \& R.~M. {Humphreys},
  221

\bibitem[{{Viotti} {et~al.}(2007){Viotti}, {Galleti}, {Gualandi}, {Montagni},
  {Polcaro}, {Rossi}, \& {Norci}}]{Viotti2007}
{Viotti}, R.~F., {Galleti}, S., {Gualandi}, R., {et~al.} 2007, \aap, 464, L53

\bibitem[{{Viotti} {et~al.}(2006){Viotti}, {Rossi}, {Polcaro}, {Montagni},
  {Gualandi}, \& {Norci}}]{Viotti2006}
{Viotti}, R.~F., {Rossi}, C., {Polcaro}, V.~F., {et~al.} 2006, \aap, 458, 225

\bibitem[{{Walborn} {et~al.}(2017){Walborn}, {Gamen}, {Morrell}, {Barb{\'a}},
  {Fern{\'a}ndez Laj{\'u}s}, \& {Angeloni}}]{Walborn2017}
{Walborn}, N.~R., {Gamen}, R.~C., {Morrell}, N.~I., {et~al.} 2017, \aj, 154, 15

\bibitem[{{Weis}(2011)}]{Weis2011}
{Weis}, K. 2011, in IAU Symposium, Vol. 272, Active OB Stars: Structure,
  Evolution, Mass Loss, and Critical Limits, ed. C.~{Neiner}, G.~{Wade},
  G.~{Meynet}, \& G.~{Peters}, 372--377

\bibitem[{{Weis}(2012)}]{Weis2012ASPC}
{Weis}, K. 2012, in Astronomical Society of the Pacific Conference Series, Vol.
  465, Proceedings of a Scientific Meeting in Honor of Anthony F. J. Moffat,
  ed. L.~{Drissen}, C.~{Robert}, N.~{St-Louis}, \& A.~F.~J. {Moffat}, 213

\bibitem[{{Wolf}(1989)}]{Wolf1989}
{Wolf}, B. 1989, \aap, 217, 87

\bibitem[{{Zharova} {et~al.}(2011){Zharova}, {Goranskij}, {Sholukhova}, \&
  {Fabrika}}]{Zharova2011}
{Zharova}, A., {Goranskij}, V., {Sholukhova}, O.~N., \& {Fabrika}, S.~N. 2011,
  Peremennye Zvezdy Prilozhenie, 11, 11

\end{thebibliography}

\end{document}